\documentclass[final,5p,times,twocolumn]{elsarticle}

\usepackage{amssymb}
\usepackage{amsmath}

\usepackage{array,multirow,makecell} 
\usepackage{booktabs} 
\usepackage{threeparttable}

\begin{document}

\begin{frontmatter}

\title{A prototype gas cell for the stopping, extraction and neutralization of radioactive nuclei from the SPIRAL2 Super Separator Spectrometer (S$^3$)}

\author[IJCLab]{W.~Dong\fnref{Wenling}} 
\author[IJCLab]{V.~Manea\corref{ref1}} 
\author[Leuven]{R.~Ferrer} 
\author[IJCLab]{S.~Franchoo} 
\author[GANIL]{S.~Geldhof} 
\author[Leuven]{F.~Ivandikov} 
\author[GANIL]{N.~Lecesne} 
\author[IJCLab]{D.~Lunney} 
\author[IJCLab]{V.~Marchand} 
\author[IJCLab]{E.~Minaya Ramirez} 
\author[IJCLab]{E.~Morin} 
\author[GSI,HIM]{S.~Raeder} 

\address[IJCLab]{Université Paris-Saclay, CNRS/IN2P3, IJCLab, Orsay 91405, France}
\address[Leuven]{KU Leuven, Instituut voor Kern- en Stralingsfysica, Leuven B-3001, Belgium}
\address[GANIL]{GANIL, CEA/DRF-CNRS/IN2P3, Caen 14076, France}
\address[GSI]{GSI Helmholtzzentrum f\"{u}r Schwerionenforschung GmbH, Darmstadt 64291, Germany}
\address[HIM]{Helmholtz-Institut Mainz, Mainz 55099, Germany}

\fntext[Wenling]{This work contains part of the PhD thesis work of Wenling Dong}
\cortext[ref1]{Corresponding author: vladimir.manea@ijclab.in2p3.fr}



\begin{abstract}
We present the design and simulation of a prototype gas cell for in-gas-jet laser-ionization and spectroscopy studies using the low energy branch of the SPIRAL2-S$^3$ radioactive-ion-beam facility. The prototype aims to demonstrate the possibility to reduce the extraction time of radioactive ions from the gas cell, while implementing a controlled neutralization mechanism, necessary for laser-spectroscopy studies. Different simulation methods of ion processes in gas are comparatively discussed. Design considerations and detailed simulations of the ion extraction time and efficiency are presented. A study of the dynamics of electrons obtained in the gas cell by ionization is also performed to assess the achievable electron densities. 
\end{abstract}


%


\end{frontmatter}



\section*{Introduction}
\label{intro}

The evolution of nuclear structure with neutron-proton imbalance is one of the major topics of contemporary experimental nuclear research. To enable its study, radioactive-ion-beam (RIB) facilities are required to produce ever more exotic nuclei. Despite great advances in radioisotope production, their typical short lifetimes, low yields and significant background of contaminant isotopes pose considerable experimental challenges. The most exotic species are produced at high-energy, in-flight facilities so that measurements requiring low-energy beams must use a gas-filled stopping cell \cite{Blumenfeld13}.

Following decades of developments, resonance laser ionization has been adopted as a highly selective ion-source method for the production of low-energy RIBs \cite{Fedossev2012}. With it, in-source laser spectroscopy \cite{Marsh2013,Kudryavtsev2013} has evolved as an attractive method to determine nuclear properties such as spins, moments and charge radii \cite{Yang_2023}. Although the environment (temperature and pressure) close to production prevents them from achieving similar spectral resolution as, for example, collinear laser spectroscopy, in-source methods, especially if combined with selective detection, tend to be more efficient and sensitive.

This work is dedicated to in-gas laser ionization and spectroscopy (IGLIS) \cite{Kudryavtsev2013}, which is a variant of in-source laser spectroscopy applied in gas stopping cells.  Since its inception in the late 1980s at LISOL in Louvain-la-Neuve, IGLIS has undergone significant advancements \cite{van1992laser,kudryavtsev2001gas,Kudryavtsev2013,Rafael2017,Zadvornaya2018} including the development of the in-gas-jet laser-spectroscopy method, which aims to probe the radioactive atoms in the supersonic gas jet produced by the carrier buffer gas expanding out of the gas cell through the exit nozzle. The significantly lower temperature and pressure of the jet with respect to the stopping cell environment leads to an order-of-magnitude improvement in the spectral resolution, while preserving the high-efficiency inherent to the in-source technique. The IGLIS technique is applied nowadays to several RIB facilities, such as KISS \cite{HIRAYAMA201711RIKEN} and PALIS \cite{zagrebaev2014gasRIKEN} at RIKEN RIBF (Japan), the RADRIS/JETRIS setups at GSI (Germany)  \cite{raeder2023RADRIS,JetRIS2024}, as well as the future facilities MARA-LEB at University of Jyväskylä (Finland) \cite{Papadakis2016MARA} and S$^{3}$-LEB at SPIRAL2-GANIL (France) \cite{Ajayakumar2023}.

The application of the IGLIS technique (and its in-jet method) at the next-generation facilities faces however a number of challenges in the extraction and the neutralization of the RIBs. First of all, to ensure the neutralization of the radioactive products, the IGLIS technique has been applied almost exclusively in gas stoppers with flow-based transport (no electrical field), which allow neutralization by recombination with free electrons produced by the energetic ion beams (either primary, or secondary) in the gas. This inherently slow transport is a severe limitation for the study of short-lived species. Second of all, this neutralization mechanism is highly dependent on the intensity of the beams entering the gas catcher, which define the density of the induced electron cloud \cite{Ferrer2013}. At LISOL, the primary cyclotron beam was traversing the gas cell and produced a significant ionization of the buffer gas \cite{kudryavtsev2001gas, Facina2004}. In the new IGLIS setups such as S$^3$-LEB, the reaction takes place outside the gas cell and only the secondary beams enter the gas \cite{Dechery2016}. Much less intense, especially for heavy-element fusion-evaporation reactions, these secondary beams might provide insufficient electron density for their efficient recombination. Finally, the whole neutralization mechanism collapses if one aims to reduce the extraction time of the RIBs by employing an electrical field, which would quickly remove any electrons from the beam stopping volume. Alternative neutralization mechanisms are therefore required \cite{JetRIS2024}. 

To address these challenges in the framework of the S$^3$-LEB experiment, a new project called Fast Radioactive Ion Extraction and Neutralization Device for S$^3$ (FRIENDS$^3$) has been launched \cite{FRIENDS3}. Within this project, a prototype gas cell and a new experimental setup for its characterization have been designed. The operational principle of the prototype gas cell is to extract the radioactive ions from the stopping volume (where the gas is almost stagnant) by an electrical field and to induce their neutralization much closer to the gas-cell exit, where transport by pure gas flow is fast enough. In this article, we report the simulation and design of this prototype. We discuss comparatively different methods to simulate the extraction of the ions in the combined action of the gas flow and of the electrical field. In connection to the FRIENDS$^3$ design of the gas cell, we also simulate the dynamics of free electrons/ions produced by the ionization of the buffer gas, taking into account their diffusion, migration in the electrical field and convection in the gas flow. For comparison, we present a simulation of the neutralization phenomenon in the existing S$^3$-LEB gas cell, considering the three-dimensional distribution of charge densities.

The article is organized as follows. Section 1 presents in more detail the operational principles and the limitations of the current S$^3$-LEB gas cell. Section 2 presents the objectives of the FRIENDS$^3$ project and the design concepts leading to the prototype gas cell. Section 3 gives an overview of the physical phenomena that must be accounted for in gas-cell simulations. Finally the simulations and optimizations of the FRIENDS$^3$ gas-cell prototype (with respect to the extraction time and efficiency), as well as the simulations of neutralization and charge dynamics in the gas, are presented in Sections 4 and 5, respectively. 

\section{Principle of the S$^3$-LEB gas cell and its limitations}
\label{limitation}

In the IGLIS approach implemented at S$^3$-LEB, nuclear-reaction products from the S$^3$ spectrometer of SPIRAL2 are stopped, thermalized, neutralized and transported within an argon buffer-gas cell \cite{Kudryavtsev2016} towards the exit aperture. The neutralized atoms are then reionized using element-selective, step-wise resonant laser excitation, either inside the gas cell near the exit aperture or within the homogeneous, cold supersonic jet formed by a de Laval nozzle \cite{Kudryavtsev2013}. The photoions are extracted from the jet area by a series of RFQ ion guides and delivered to a multi-reflection time-of-flight mass spectrometer (MR-TOF MS \cite{Chauveau16}), where the ions can be separated by mass before detection \cite{Ajayakumar2023}. 

Upon entering the gas cell, a few percent of the energetic ions will neutralize by direct charge-exchange with the argon buffer gas \cite{Lautenschlager2016}. Following the thermalization of the surviving ions, however, charge-exchange will no longer be energetically possible. The remaining neutralization occurs by recombination with free electrons resulting from the ionization of the buffer gas in the stopping process \cite{kudryavtsev2001gas, Facina2004, Moore2010}. Two processes are possible. Firstly, three-body recombination, where an electron is captured in a high orbit of the ion and a collision with a buffer-gas atom stabilizes the neutral compound. Secondly, dissociative recombination, where the ion first forms an ionic molecule with a gas component, which then captures an electron and breaks, releasing the radioactive species in atomic form. The rate of both processes depends significantly on the density of electrons present in the gas, which in turn depends on the specifics of the ion-stopping process (intensity and energy of the ion beam, gas pressure). Furthermore, maintaining an electron density sufficient for gas-phase neutralization requires the absence of any electrical field, because the high electron mobility would lead to their immediate removal in an opposite direction to the ion transport, before recombination can take place.  

The current version of the S$^3$-LEB gas cell \cite{Kudryavtsev2016} is thus designed to facilitate recombination with electrons. Ions are guided from the stopping region to the exit only by the laminar argon gas flow. The two additional objectives which are taken into account in its design are to achieve an extraction time as low as possible and to achieve good stopping and extraction efficiency. In the absence of any electrical field which could reduce extraction time, the S$^3$-LEB gas cell is conceived to have the minimal volume possible for stopping the ion beam, with a depth of only 30~mm. This minimal stopping volume is necessary to accommodate the large S$^3$ beam (about 20~mm FWHM in converging mode \cite{Dechery2015}), which leads to an extraction time of about 600~ms for a 1~mm nozzle throat diameter, and roughly half that time for a 1.5~mm diameter throat \cite{Mogelevskiy2013a,Mogelevskiy2013b,Kudryavtsev2016, Verlinde2021}. The small volume can lead to losses of the radioactive beam, either due to a too long stopping range (larger than the gas-cell depth), or due to diffusion to the walls of the gas cell during the long extraction time. To mitigate both aspects, the S$^3$-LEB gas cell is designed to be operated at 500~mbar, which increases stopping power and significantly reduces the diffusion phenomenon.   

The resulting design of the gas cell is adequate for the first experiments proposed for the S$^3$ spectrometer, however a number of its performance parameters remain quite tightly constrained. First of all, the extraction time and efficiency depend in a critical way on the range of the beam in the gas cell, as shown in \cite{Kudryavtsev2016}, both parameters being significantly affected if the beam is stopped close to the wall. In the case of very asymmetric reactions resulting in slow reaction products, if the entrance-window thickness cannot be  sufficiently reduced, the gas pressure will have to be reduced instead, to achieve the desired stopping range. This will potentially increase diffusion losses. Furthermore, for reactions with relatively low production rates, the achievable electron densities for neutralization can be quite low, leading to insufficient neutralization, even in simple estimates that do not account for the spatial distribution of the energy deposited by the ion beam in the gas cell, or for electron diffusion \cite{Ferrer2013}. Finally, for short-lived isotopes or isomers, the long extraction time can lead to significant losses by decay. Reducing the extraction time even further by simply reducing the volume of the gas cell is not easily achievable considering the S$^3$ beam size and the requirement to stop the beam far from the gas-cell walls. Further increasing the exit hole size, on the other hand, comes at a significant cost in gas consumption and required pumping capacity for achieving a homogeneous, high Mach-number jet. 

All these limitations and critical constraints of the S$^3$-LEB gas cell motivate exploring an alternative approach to its design, as will be discussed in the next section.
 
\section{Conceptual design of the FRIENDS$^{3}$ gas cell}
\label{friends}
 
The total extraction time of the S$^3$-LEB gas cell is about 600-ms \cite{Mogelevskiy2013a,Mogelevskiy2013b,Kudryavtsev2016, Verlinde2021}. The largest part of this time ($\approx$ 400~ms) is necessary solely for extracting the radioactive species from the stopping volume, where the gas is almost stagnant. On the other hand, the evacuation time of the ions from the final extraction channel of the gas cell is only a few tens of ms, due to the higher gas-flow velocity. The main approach of the FRIENDS$^3$ project is thus to separate stopping and neutralization: use an electrical field to quickly extract the radioactive species still in ionic from the large stopping volume with slow gas flow, then transfer them to the fast gas-flow region close to the exit, where they can be neutralized in absence of electrical fields.  

It is worth mentioning that the JetRIS project at GSI has already made a step in this direction \cite{Raeder2020} by combining ion extraction from the stopping volume by electrical field with collection/neutralization onto a heated filament, followed by re-evaporation, the latter approach being well established in the RADRIS technique for the gas-cell laser spectroscopy of very heavy elements \cite{Laatiaoui16}. The method has already been applied on-line to perform the gas-jet laser spectroscopy of $^{254}$No \cite{JetRIS2024}, but so far the efficiency of the technique has been limited, with the causes still under investigation and potentially related to ion-transport losses or space-charge effects. Furthermore, the evaporation temperature from the filament (or conversely the release time) is element-dependent and can be a limiting factor for high-melting-point elements.  

In this work, we aim to explore an alternative approach which also employs the fast extraction of ions by an electrical field from the stopping volume, but which does not rely on filament collection in the exit channel of the gas cell. Instead, once in the exit channel, the ions would neutralize either by recombination with free electrons, or by another process in the gas phase. Neutralization in the channel volume would make the process less dependent on ion transport to a specific collection point. As working hypothesis, we consider recombination with electrons produced by the ionization of the buffer gas, either by irradiation with decay particles (such as suggested in \cite{Ferrer2013}), or by electrical discharge. 

To optimize the geometry and evaluate the performance of the design, three parameters are quantified by simulations which will be presented in the following sections: extraction efficiency $\epsilon_\text{et}$ (ratio of extracted ions to ions stopped in the gas cell), extraction time $t_\text{ext}$, and the time available in the exit channel for ion neutralization $t_\text{n}$. The design optimization aims to minimize $t_\text{ext}$, maximize $\epsilon_\text{ext}$, and ensure sufficient $t_\text{n}$ for the neutralization to take place. These objectives often involve trade-offs, for example a longer neutralization channel is expected to enhance neutralization efficiency by providing a longer interaction time, but increases ion residence time in the gas cell and reduces efficiency due to diffusion losses to the gas-cell walls. 

In addition to the optimization of the gas-cell extraction efficiency and time, we also perform in this work a first step in studying the dynamics of electrons generated by gas ionization in the exit channel of the gas cell. For this study, the figure of merit is the achieved equilibrium density, required to ensure efficient recombination before the ions exit the gas cell. The simulations aim especially to quantify the effect of diffusion, migration in the electrical field and convection on the equilibrium density. 

Since the final product of this development should be a gas cell to be installed at the S$^3$ focal plane or in the S$^3$-LEB vacuum chamber, certain spatial constraints apply, as detailed below. The total gas-cell length and diameter is constrained by the available space in the S$^3$-LEB vacuum chamber. Furthermore, the gas cell's inlet and exit are designed to match the S$^3$-LEB dimensions. The S$^3$ beam enters through a 74-mm-diameter window optimized to the S$^3$ beam in convergent mode \cite{Dechery2016}. The gas cell's exit aperture is set to the standard 1-mm diameter, a value determined by the vacuum system's pumping capacity and an economical approach to gas consumption. 

Second of all, as in the case of S$^3$-LEB and JetRIS \cite{JetRIS2024}, argon is used as a buffer gas, because of its larger stopping power (important for achieving a compact gas cell), lower diffusion of the stopped ions and generated electrons (required for minimizing losses in the neutralization channel), lower ionization potential and larger recombination coefficient (which favor neutralization). 

Third of all, for the prototype presented in this work, a simplified geometry is chosen (compared to the S$^3$-LEB gas cell), to ensure computational efficiency during the optimization process (for example, the use of cylindrical symmetry in the simulations), and to minimize machining complexity. Furthermore, the current prototype uses a pure DC guiding field despite its known efficiency limitations, compared to the use of RF guiding in a helium environment \cite{Karvonen_2008}. However, we consider that the type of electrical-field guiding is a secondary issue, provided that a satisfactory transport efficiency to the neutralization channel can be achieved. The use of DC fields avoids the need for RF circuits in the gas environment, which would make it more challenging to maintain high purity in the gas, without opting for a cryogenic gas cell \cite{DROESE2014He_cell}. 

\section{Theoretical aspects of gas-cell simulations}
\label{phenmena}

\subsection{Simulation of ion transport}

The transport of ions in the gas cell takes place under the combined action of the electrical force and of the collisions with the buffer-gas atoms. The former depends on the electrostatic field, which can be determined at every point along the ion trajectory by solving Poisson's equation with boundary conditions defined by the gas-cell/electrode geometry and applied voltages. Concerning ion-neutral collisions, while individual modeling is feasible in low-pressure regimes ($<$ 1 mbar) their collision frequency at the higher pressures targeted for this design (50–500~mbar) make it computationally prohibitive and one must opt for different numerical solutions.

The simplest one is to introduce a viscous-damping model, where the effect of the particle moving in the gas is modeled by a drag force. The drag force can be described as:
\begin{equation}
    \mathbf{F}_\text{drag} = -m \delta (\mathbf{v - u}),
    \label{eq:F_drag}
\end{equation}
where \(m\) is the ion mass, \(\delta\) the damping coefficient, \(\mathbf{v}\) the ion velocity, and \(\mathbf{u}\) the gas flow velocity. In a stationary gas (\(\mathbf{u} = 0\)), the drag force simplifies to \(\mathbf{F}_\text{drag} = -m\delta \mathbf{v}\). At equilibrium, the drag force balances the electric force (\(\mathbf{F}_\text{e} = q\mathbf{E}\)), yielding the drift velocity:
\begin{equation}
    \mathbf{v}_\text{drift} = \frac{q}{m\delta} \mathbf{E} = K \mathbf{E},
    \label{eq:vdr}
\end{equation}
where \(K\) is the ion mobility, expressed as:
\begin{equation}
    K = K_0 \frac{T/T_N}{P/P_N},
    \label{eq:k0}
\end{equation}
with $K_0$ being the reduced mobility specific to the ion species and buffer gas, $T$ and $P$ the gas temperature and pressure, and \(T_N = 273.15~\text{K}\), \(P_N = 1~\text{atm}\) the standard conditions \cite{mcdaniel1973mobility}. It is clear from Eq.~(\ref{eq:F_drag}) that determining the viscous-damping force requires computing the gas-velocity, pressure and temperature fields in the gas cell. 

The viscous-damping force allows describing the deterministic component of the ion trajectories under the combined action of the electrical field and gas flow. Nevertheless, the diffusion phenomenon, which is important especially in lower pressures, is not captured. Since direct modeling of ion-neutral collisions in not possible as mentioned above, a statistical simulation of the diffusion process is necessary. 

We therefore opt for a simulation approach combining the COMSOL finite element analysis (FEA) software \cite{COMSOL5.6} and SIMION \cite{2000Dahl}. Gas flow simulations were conducted using the COMSOL Laminar Flow Interface, solving the Navier-Stokes equations for compressible laminar flow under low Mach number conditions ($M<0.3$). The ion transport was then simulated with SIMION using the Statistical Diffusion Simulation (SDS) model \cite{Appelhans2005}. For these simulations, the gas velocity fields computed with COMSOL were exported and then mapped to a grid-type input file used by SIMION, while pressure and temperature were treated as uniform fields compatible with the COMSOL simulation conditions. 

The SDS model of SIMION implements the drag force of Eq.~(\ref{eq:F_drag}) and treats diffusion as a statistical process, where for each trajectory integration step it computes a displacement $\lambda_N$ proportional to the square root of the number of collisions ($\sqrt{N}$) in the integration time \cite{Appelhans2005}. The SDS model uses pre-computed scattering radii from Monte Carlo simulations, accounting for various ion-gas mass ratios. At each integration step, the scattering radius is interpolated, and the physical scattering radius $\lambda_{N_1}$ is calculated using the mean free path $\lambda_0(P, T)$ derived from the gas pressure and temperature. The ion trajectory is then adjusted with a random displacement based on the computed scattering radius.

In the optimization phase of the gas-cell geometry, COMSOL-only simulations were also performed using the Electrostatic interface for solving the Poisson equation and the Charged Particle Tracing (CPT) module for ion-trajectory calculation. For the latter, the drag force was introduced as a user-defined force, in addition to the electrical one. Although overestimating efficiency these simulations allowed for a comparative study of the effects of different voltages and geometries on the idealized ion trajectories. SIMION and the SDS module were then used to characterize with diffusion included an optimized geometry and voltage set.

A second approach to simulate ion transport in the gas cell will be presented in the next section together with the study of electron densities.

\subsection{Simulation of ion and electron densities}

While the simulation of individual ion trajectories is ideal for studying the gas-cell transport of ion and electron beams, a treatment based on particle densities is more appropriate for studying the generation and evolution of electron clouds by gas ionization, or for modeling the processes taking place between ion, electrons and other species in the gas phase. Concerning the pure charge dynamics, the Plasma module of COMSOL includes the relevant equations for the electron and ion densities in the Drift-Diffusion and Heavy-Species Transport interfaces, respectively. They can be written as:

\begin{equation}
     \frac{\partial n_k}{\partial t} + \mathbf{\nabla}\cdot (z_k n_k \mu_k \mathbf{E}  - D_k\mathbf{\nabla} n_k)  = R_k - (\mathbf{u} \cdot\mathbf{\nabla})n_k,
    \label{eq:n_e_COMSOL} 
\end{equation}
where $n_k$ is the particle density and the index $k$ can be either $e$ for electrons or $i$ for ions. In Eq.~(\ref{eq:n_e_COMSOL}), the first term represents the density time evolution. The second term is the divergence of the flux, which comprises two primary transport mechanisms: migration ($z_k n_k \mu_k \mathbf{E}$) driven by the electric field, and diffusion ($ - D_k\mathbf{\nabla} n_k$) due to the concentration gradient. $\mu_k$ and $D_k$ are the mobility and diffusion coefficients, while $z_k$ represents the charge sign, $+1$ for ions and $-1$ for electrons. On the right-hand side, the source term $R_k$ quantifies the rate of electron/ion production or loss, summing over the contributions of all reactions or sources, such as ionization and recombination. The convection term $(\mathbf{u} \cdot\mathbf{\nabla})n_k$ describes the influence of gas flow, where $\mathbf{u}$ represents the gas velocity field. One should note that in COMSOL, ions and neutrals are modeled not by densities but by mole fractions, but the corresponding equations can be reduced in our case to the form of Eq.~(\ref{eq:n_e_COMSOL}).

With Eq.~(\ref{eq:n_e_COMSOL}), the Plasma module of COMSOL can be used to simulate two types of phenomena. On the one hand, for an ion species, by setting the source term $R_i$ to zero and defining an initial ion density distribution, it is possible to describe ion propagation in the gas cell under the combined action of the electrical field and gas flow, while accounting for diffusion. This provides a second method for simulating ion transport in the gas cell and for determining the extraction time and efficiency. Like in the case of the SDS model in SIMION, this approach accounts for the effect of the electrical field, gas flow and diffusion. On the other hand, in the case of electrons, by defining a production rate $R_e$ (for example, by background-gas ionization) and by coupling other reactions or processes (such as recombination with the ions of the background gas, or inelastic scattering on the background gas) it is possible to simulate the dynamics of the electron cloud by considering all relevant phenomena. Both applications of the Plasma module will be explored in this work.

It is worth noting that in the design phase of the S$^3$-LEB gas cell, the extraction efficiency and time of the design were simulated using COMSOL and the Transport of Diluted Species (TDS) interface \cite{Mogelevskiy2013a,Mogelevskiy2013b,Kudryavtsev2016}. The equations solved by the TDS module are similar to Eq.~(\ref{eq:n_e_COMSOL}), the difference being that, instead of mole fractions, the species were modeled by particle concentrations. Results of this type of simulation are reported for example in \cite{Kudryavtsev2016, Verlinde2021}. Electrical fields can be included in the TDS physics interface, but for the S$^3$-LEB gas cell they were not considered. At the beginning of Section~\ref{Electron}, we present a new series of simulations of the S$^3$-LEB gas cell using the TDS module. What is important to emphasize is that, while they are not using the Plasma module as the other simulations of the same section, they are for all practical purposes also solving Eq.~(\ref{eq:n_e_COMSOL}). 

\section{Design and simulation of FRIENDS$^3$ gas cell}
\label{fr3gc}

\subsection{Simulation parameters}
\label{params}

The common parameters of the presented simulations are the dynamic viscosity of the gas (important together with the gas density in the laminar flow simulations) and the ion/electron mobilities, responsible for the drift phenomenon in all the discussed simulation methods (COMSOL with the CPT or the Plasma module, or SIMION with the SDS module). For the simulation of the diffusion phenomenon, different parameters are used by the Plasma or SDS modules. The former uses the diffusion coefficient $D$, which for ions is computed from the mobility using the Einstein relation $D = K_0 k_B T/e$, where $K_0$ is the reduced ion mobility, $k_B$ is the Boltzmann constant, $T$ is the absolute temperature and $e$ is the elementary charge. The SDS module does not simulate diffusion explicitly, but rather uses the ion and gas diameters in order to estimate the mean free path, determining the number of collisions per integration step. The ion diameter is estimated based on systematic trends considering the ion mass and its reduced mobility. For the simulations of in-gas electron production and dynamics with the Plasma module, the total recombination rate of electrons with ions of the buffer gas is an  important parameter which influences the equilibrium density. In addition, the electron diffusion coefficient and mobility have to be considered.

Using the combined COMSOL and SIMION simulation methodology described in the previous section, ion trajectories were modeled with all relevant physical phenomena accounted for. Simulations were performed using $^{133}$Cs$^+$ ions in argon which is taken as the buffer gas for all simulations. The choice for $^{133}$Cs$^+$  is due to its well known reduced ion mobility, $K_0$($^{133}$Cs$^+$) = 2.1 cm$^2$V$^{-1}$s$^{-1}$ \cite{Appelhans2005}, and its planned use in subsequent experimental validations. The dynamic viscosity for argon was taken to be $2.26 \times 10^{-5}$~Pa~s \cite{Huber2011}. The Einstein relation led to a diffusion coefficient for $^{133}$Cs$^+$ in argon of 0.053~cm$^2/$s (at a pressure of 1~atm).

\begin{table}
\small
\centering
\caption{Key design parameters of the FRIENDS$^3$ gas cell in its two simulated versions.}
\vspace*{1mm}
\begin{tabular*}{\linewidth}{@{\extracolsep{\fill}}lll}
\toprule
\multirow{2}{*}{Component }                    & \multicolumn{2}{c}{Design version }   \\
&Initial&Final\\ \midrule
Buffer chamber               &                                           \\
\hspace{5mm}Length          & 233.0 mm             & 235.0 mm                                                \\
\hspace{5mm}Diameter        &120.0 mm             & 200.0 mm                                                 \\
\hspace{5mm}Exit hole diameter   & 1.0 mm        & 1.0 mm  
\\ 
\hspace{5mm}Window diameter & 74.0 mm& 74.0 mm                                         \\
\hspace{5mm}Window depth & 30.0 mm & 30.0 mm                                         \\

\hspace{5mm}Disk internal diameter  & 80.0 mm      & 80.0 mm                                               \\\midrule
DC-cage                     &                                                       \\
\hspace{5mm}Cage length    & 70.0 mm  & 72.0 mm                                               \\
\hspace{5mm}Cage inner diameter   & 78.0 mm   & 110.0 mm                                               \\
\hspace{5mm}Cage outer diameter  & 100.0 mm    & 158.0 mm                                               \\
\hspace{5mm}Number of rings    & 5           & 5                                                    \\
\hspace{5mm}Electrode thickness    & 2.0 mm           & 2.0 mm                                                \\
\hspace{5mm}Electrode spacing    & 8.0 mm        & 8.0 mm                                                \\ \midrule
DC-funnel                     &                                                       \\
\hspace{5mm}Funnel length    & 46.0 mm  & 48.0 mm                                               \\
\hspace{5mm}Entrance inner diameter & 78.0 mm     & 110.0 mm                                               \\
\hspace{5mm}Exit inner diameter  & 18.0 mm        & 18.0 mm                                                \\
\hspace{5mm}Number of rings      & 8         & 5                                                    \\
\hspace{5mm}Electrode thickness   & 2.0 mm           & 2.0 mm                                                \\
\hspace{5mm}Electrode spacing     & 2.0 mm       & 2.0 mm   
           \\ \midrule
Neutralization channel     &        &                                                       \\
\hspace{5mm}Cylindrical part length         & 50.0 mm              & 50.0 mm                                               \\
\hspace{5mm}Cylindrical part diameter          & 18.0 mm           & 16.0 mm                                                 \\                                           
\hspace{5mm}Convergent part length  & 20.0 mm &       20.0 mm                                       \\
\hspace{5mm}Segment length                 & 4.0 mm  & -                                              \\ 
\hspace{5mm}Segment spacing                 & 1.0 mm  & -                                              \\ 
\bottomrule
\end{tabular*}
\begin{tablenotes}\footnotesize
\item ``--'' indicates elements not included in the final design.
\end{tablenotes}
\label{tab:gas_cell}
\end{table}

Ions were released in our simulations following a 3D Gaussian spatial distribution with a depth dispersion ($\sigma_z$) of 5 mm, x/y dispersion ($\sigma_{x,y}$) of 10 mm, and a release depth of 50 mm to approximately model the stopped online S$^3$ beam in converging mode \cite{Dechery2016}. The recorded number of particles reaching the exit and their arrival times enabled the determination of extraction efficiency ($\epsilon_\text{ext}$) and extraction time ($t_\text{ext}$). Additionally, the average time of flight within the field-free region ($t_\text{n}$) provided an estimate of the time available for ion neutralization. This approach facilitated iterative optimization of the design to achieve the best performance quantified by $\epsilon_\text{ext}$, $t_\text{ext}$ and $t_\text{n}$. We note that, although some of the simulations shown in the following correspond to different stagnation pressures in the gas cell, the initial ion distribution is not changed, although in a real situation both the stopping depth and the straggling of the beam would change with pressure. Nevertheless, a more realistic approach would complicate the study, because one would have to optimize for each pressure the ion-transport voltages. We decided for the sake of consistency to use the same ion distribution, with a depth dispersion comparable to the straggling obtained at 100-200~mbar (which is a worst-case scenario). Furthermore, in a real case the use of energy degraders would allow to compensate a change in pressure.

\begin{figure*}
    \centering    \includegraphics[width=0.6\linewidth]{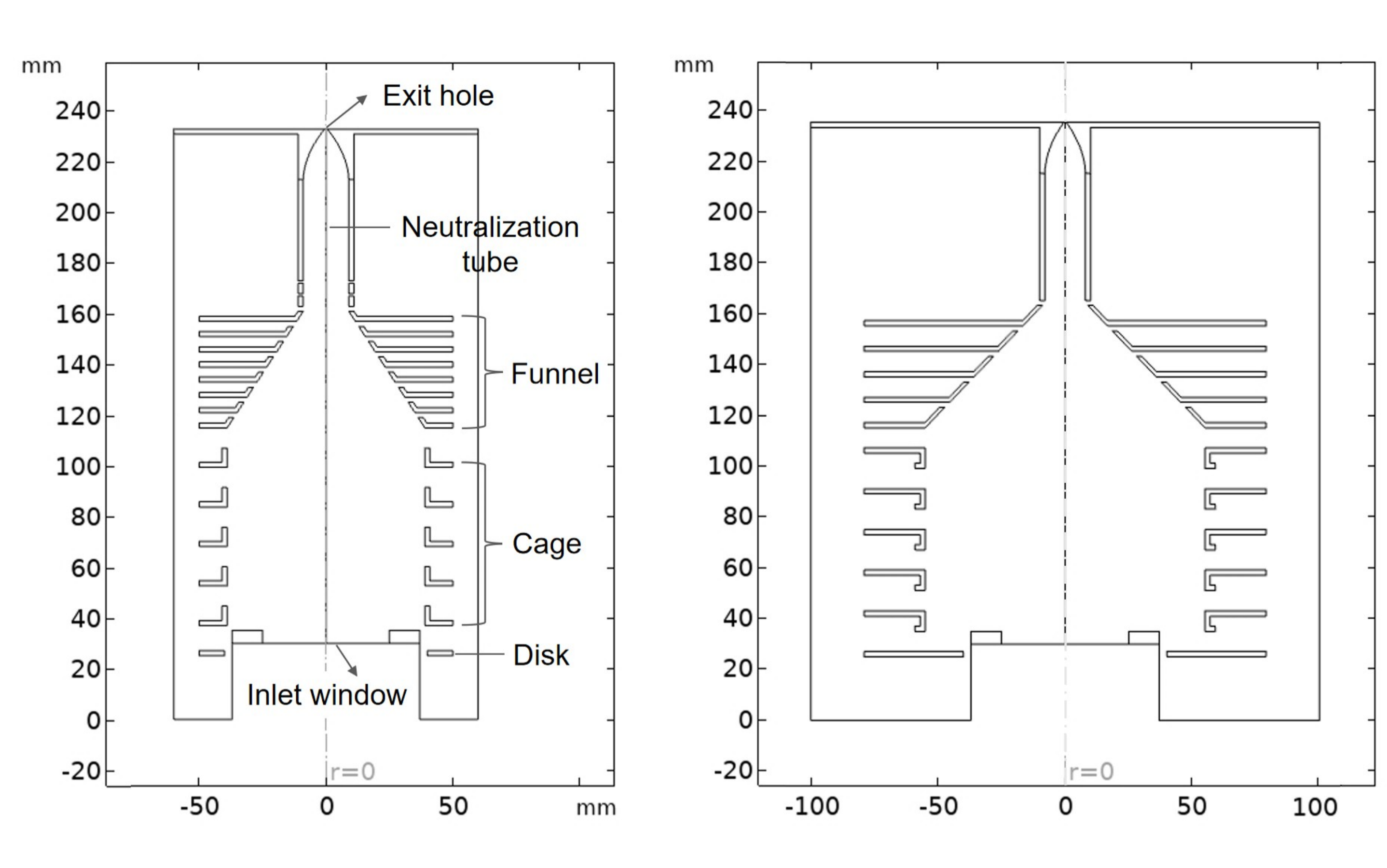}
    \caption{Schematic representation of the initial (left) and final (right) gas-cell designs. The initial design features a more compact geometry, a larger number of funnel electrodes, and a segmented neutralization channel, whereas the final design adopts a wider and simplified geometry with fewer electrodes, taking into account  stopping-volume margins, mechanical constraints and manufacturing considerations.}
    \label{fig:geo}
\end{figure*}
 
\subsection{Geometry description}
\label{geo}

The gas-cell geometry was simulated over several iterations, of which two important stages will be detailed here. The first one is an early version of the design, with a smaller volume and a larger number of electrodes, in order to maximize the flow velocity and to obtain a better control over the shape of the electrical field in the transport optimization. This design, labeled ``initial design'' in the following, was also used to compare the different simulation methods described above and to illustrate the competition between transport efficiency and extraction time. The second one is the final design which was actually manufactured, taking into account also mechanical constraints, manufacturing cost, margin for stopping volume and distance between electrodes and the vacuum chamber.  The key design parameters of both designs are detailed in Table~\ref{tab:gas_cell}.

\begin{table}
\small 
\centering 
\begin{threeparttable}
\caption{Optimized voltage configurations for the low-voltage (LV) and high-voltage (HV) settings of the initial design, as well as the high-voltage settings of the final design.} 
\vspace*{1mm}
\begin{tabular*}{\linewidth}{@{\extracolsep{\fill}}cccccc}
     \hline
     \hline
No. & Component & \multicolumn{3}{c}{DC voltage (V)} \\
  & & Init. LV &Init. HV&Fin.\\
       \hline
1 &Window/ring & 246 &362&398\\
2 &Cage disk& 256&382&420 \\
3 &Cage 1& 260&386&432 \\ 
4 &Cage 2& 252&380&428 \\ 
5 &Cage 3& 248&376&424 \\ 
6 &Cage 4& 244&356&410 \\ 
7 &Cage 5& 240&280&390 \\ 
8 &Funnel 1& 212&220&320 \\ 
9 &Funnel 2& 180&180&260 \\ 
10 &Funnel 3& 150& 150&210\\
11 &Funnel 4& 120& 120&140\\
12 &Funnel 5& 90& 90&80\\
13 &Funnel 6&  60&  60&-\\
14 &Funnel 7& 30&  30&-\\
15 &Funnel 8 & 16&  16&-\\
16 &Channel segment 1 & 8& 8&-\\
17 &Channel segment 2 & 2& 2&-\\
18 &Channel segment 3 & 0& 0&0\\
       \hline
       \hline
 \end{tabular*}
 \begin{tablenotes}\footnotesize
\item ``--'' indicates elements not included in the final design.
\end{tablenotes}
 \label{tab_LH_V}
\end{threeparttable}
\end{table}

Figure~\ref{fig:geo} illustrates a section through the initial design (left) and final design (right), as implemented in COMSOL. 
Both feature a cylindrical chamber comprising an electric-field region and a field-free channel (``neutralization tube'' in the figure). The electric-field region includes two sets of electrodes: the cage, which comprises identical cylindrical electrodes with uniform dimensions, and the funnel, which consists of electrodes with progressively decreasing inner diameters. The exit of the funnel is connected to a long, field-free cylindrical tube that tapers gradually towards the exit hole of the gas cell, facilitating a seamless connection with the de Laval nozzle. In both the COMSOL and SIMION simulations, cylindrical symmetry was used.

In the gas-flow simulation, the gas enters the cell through the inlet window and evacuates through the small exit hole, creating stagnation pressure and velocity fields in the gas cell. The temperature of the gas is fixed at 293.15 K. The gas inlet condition uses a defined mass flow, while the outlet condition uses the expression of the choked volume flow through the exit aperture \cite{Kudryavtsev2013}. Specific stagnation pressures were achieved by adjusting the inlet flow rate. To avoid simulating the gas flow in the sonic region entering the nozzle (for which the present model would not be valid) the simulation volume was stopped 2 mm before the exit hole.

In the real gas cell, a lateral gas inlet would be used, but the cylindrical symmetry cannot represent such an inlet. Nevertheless, for the scope of the simulation (and not considering turbulence), using the window as the inlet is sufficient because it provides an accurate representation of the gas flow close to the neutralization channel, where the gas flow starts to influence the ion motion.

\subsection{Tests of the simulation methodology}
\label{valid}

The optimization of the electrode voltage configuration aimed to achieve a balance between the competing objectives of rapid ion extraction and minimized ion losses, while the neutralization channel dimensions were optimized to provide several tens of ms for ion neutralization. A first set of reference voltages for the initial design, with a lower extraction gradient in the cage, are presented in the first column of Table~\ref{tab_LH_V}) and are labeled ``low-voltage configuration'', having a maximum electrode voltage of 260 V. This configuration was used to illustrate and cross-validate the different simulation methods discussed in the previous section.  

\begin{figure*}[t]
\centering
\includegraphics[width=\linewidth]{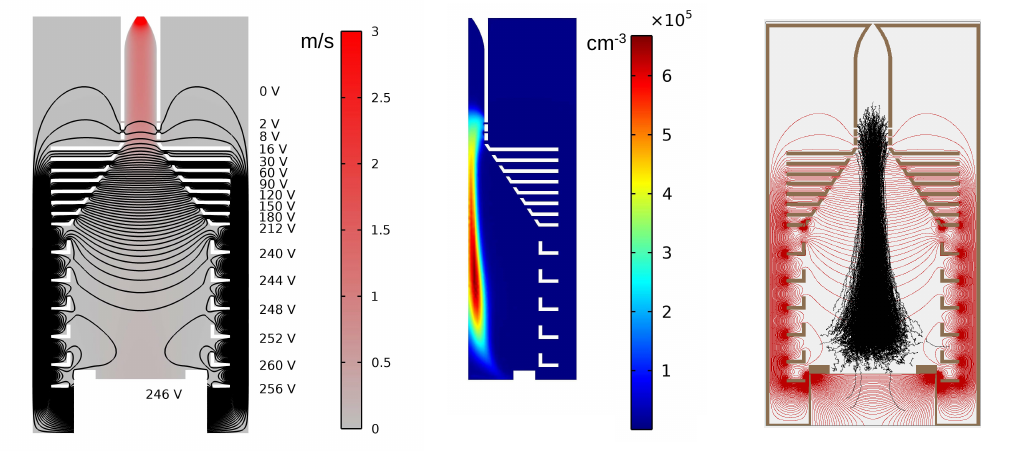}
\caption{Examples of simulations performed for the characterization of the gas-cell performance in its initial design (see Fig.~\ref{fig:geo}), at 100~mbar argon pressure and using the so-called low-voltage settings. The left panel shows the results of a combined COMSOL simulation of the gas-flow velocity (corresponding to the color legend in m/s) and of the electrical potential (electrode voltages and equipotential contours are shown). The middle panel shows a COMSOL simulation of ion transport using the Plasma module (the color legend presents the ion density in cm$^{-3}$). The right panel shows the corresponding simulation performed using SIMION (but with the individual ion trajectories shown, instead of the ion density). Both ion-transport simulations were performed with $^{133}$Cs$^+$ and diffusion is taken into account. The images in the middle and right panels are captured after 45~ms of ion transport. We note that the COMSOL simulation is performed in cylindrical coordinates. The properties of the initial ion density in the COMSOL simulation are matched to those of the initial ion distribution used in SIMION.} 
\label{COM_SIM_tra}
\end{figure*}

Figure~\ref{COM_SIM_tra} illustrates a series of representative results for these simulations. The left panel shows in color code the gas velocity and (superimposed) the equipotential lines obtained following a calculation of the electrical fields using COMSOL. The potentials applied on the electrodes in the low-voltage configuration are shown. The middle and right panels show a snapshot of the $^{133}$Cs$^+$ ion transport in the gas cell after 45~ms from the start of the simulation, using the two methods which include all relevant phenomena, namely using the Plasma module of COMSOL (middle) and the SDS model of SIMION (right). The snapshot corresponds to the moment when the first ions enter the neutralization channel. In the right panel, the equipotential contours computed using SIMION are also shown. All simulations correspond to a stagnation pressure of 100~mbar. 

\begin{figure}
\centering
\hspace*{-2mm}
\includegraphics[width=0.4\textwidth]{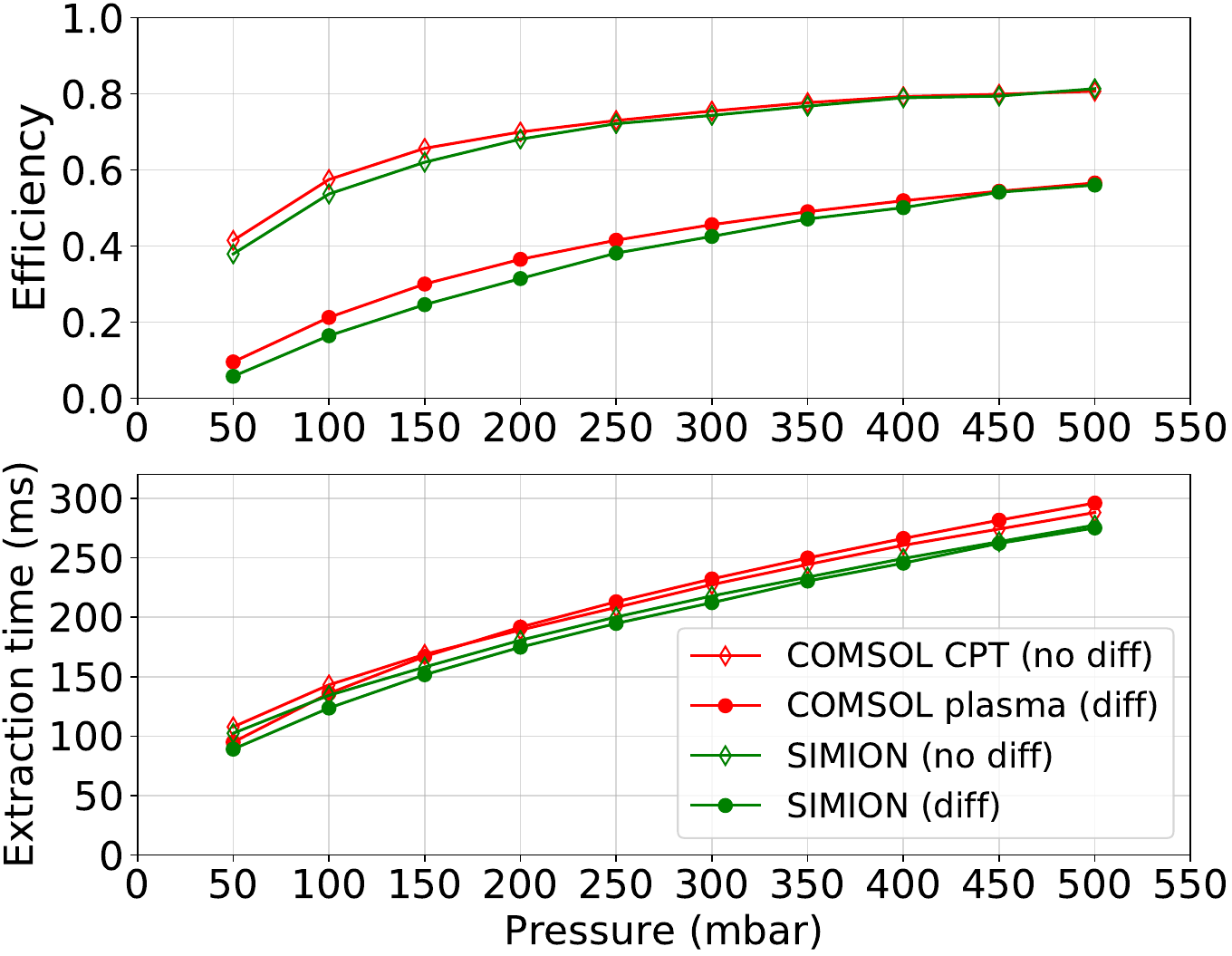}
\vspace*{0mm}
\caption{Simulated extraction efficiency (top) and average extraction time (bottom) as a function of pressure, for the initial gas-cell design with the low-voltage settings, obtained with COMSOL (red) using either the CPT module, without diffusion effects (open diamonds) or the Plasma module, with diffusion effects (closed circles), compared to results obtained using the SDS model of SIMION (green) without (open diamonds) or with (closed circles) diffusion effects.}
\label{COM_SIM_eff}
\end{figure}

The performance of the low-voltage configuration of the optimized initial design was evaluated using COMSOL and SIMION simulations across stagnation pressures ranging from 50~mbar to 500~mbar in increments of 50~mbar. The COMSOL simulations were performed using a physics-driven automatic mesh with ``finer'' size. The SIMION simulations were performed using a resolution of 10 grid units per mm. Figure~\ref{COM_SIM_eff} shows the simulated extraction efficiency and average extraction time as functions of pressure, comparing COMSOL simulations (red) using the CPT and Plasma modules with SIMION simulations (green) based on the SDS model with and without diffusion effects. Results without/with diffusion are presented with open/full symbols, respectively. 

The observed agreement between the COMSOL and SIMION simulations performed in similar conditions (in the absence or presence of diffusion) gives confidence in our simulation methodology. The results also indicate an inverse relationship between extraction efficiency and extraction time with pressure, present also in the absence of diffusion effects. According to Eq.~(\ref{eq:F_drag}), Eq.~(\ref{eq:vdr}) and Eq.~(\ref{eq:k0}), ion mobility, and consequently ion drift velocity and drag force, are inversely proportional to pressure. At higher pressures, reduced ion drift velocity results in longer extraction times. However, the increased drag forces near the window and at the entrance of the neutralization channel reduce ion losses by counteracting the electrical forces that direct ions toward the window and the channel walls. One clearly observes that diffusion is the main cause of transport losses in the gas cell, with the effect being more significant at lower pressures, where diffusion is stronger. 

\subsection{Variation of extraction-voltage gradient}
\label{performance}

To explore the possibility and limitations of further reducing the extraction time, another voltage configuration was simulated, having a stronger gradient in the gas-cell cage. The corresponding voltages are presented in the second column of Table~\ref{tab_LH_V}) and are labeled ``high-voltage configuration'', having a maximum electrode voltage of 386~V. The voltage difference between adjacent electrodes was nevertheless limited to maximum 80 V, in accordance with the Paschen curve for argon \cite{Torres_2012}, to prevent breakdown.

Figure~\ref{SIM-HV-LV} compares the simulated performance of the two voltage configurations in terms of extraction efficiency, average extraction time, and average time available in the neutralization channel as functions of the stagnation pressure. As an additional check point, the efficiency and extraction-time values are computed and represented also for the entrance of the neutralization channel, which is taken as the entrance coordinate of Funnel~7. Based on these simulation results, the following conclusions can be drawn:

\begin{figure}
\vspace*{2mm}
\centering
\includegraphics[width=0.4\textwidth]{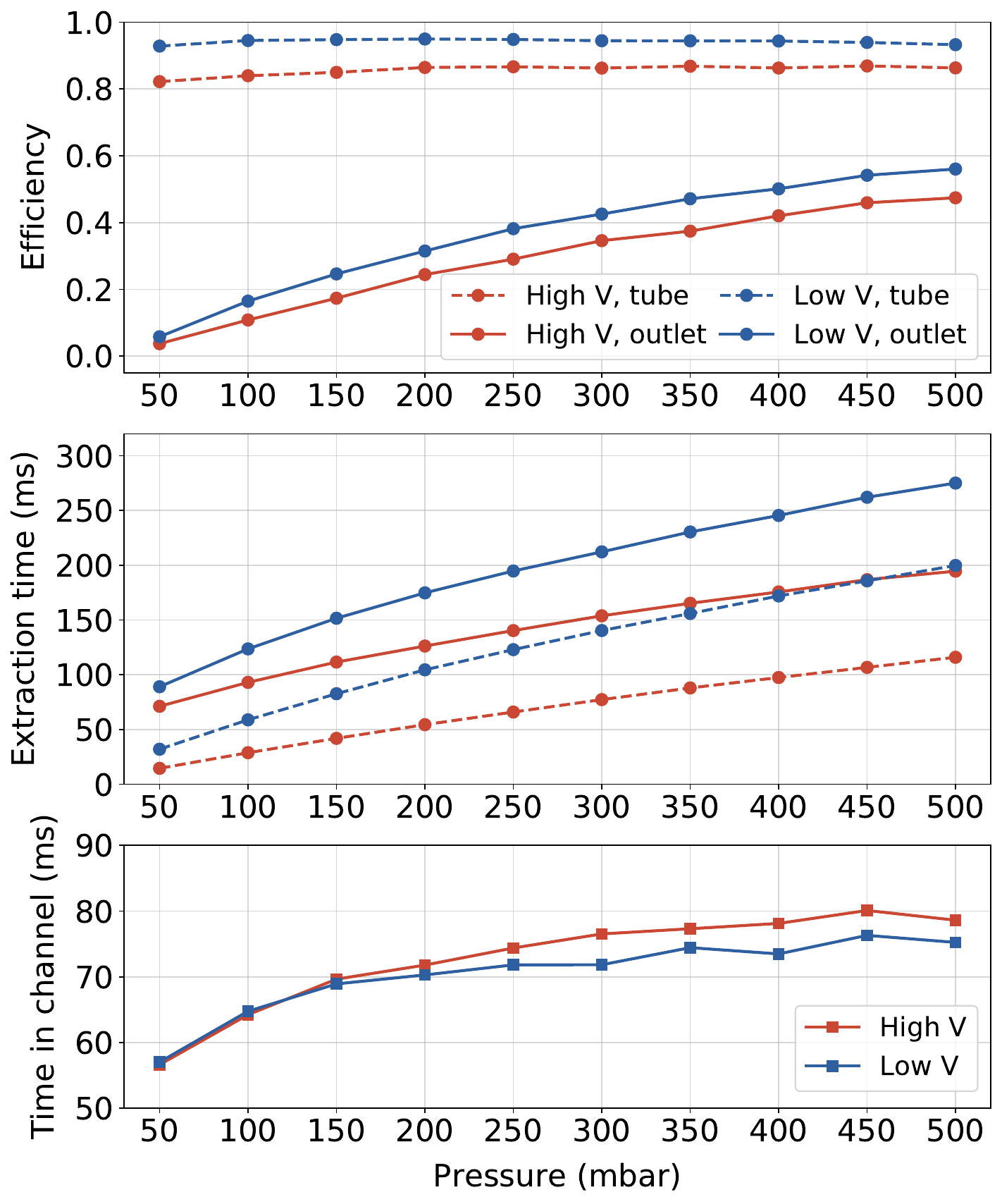}
\vspace*{1mm}
\caption[SIMION simulation of the gas cell in its initial design, as a function of pressure.]{SIMION simulation of the gas cell in its initial design as a function of pressure, comparing the low-voltage (blue) and high-voltage settings (red), see Table~\ref{tab_LH_V}. (Top) Extraction efficiency, (middle) extraction time up to the entrance of the neutralization channel (dashed lines) and up to the outlet (solid lines), and (Bottom) residence time of the ions in the neutralization channel.\label{SIM-HV-LV}}
\end{figure}

\begin{enumerate}
    \item The enhanced electric field in the high-voltage configuration enables a more rapid extraction, approximately 50~ms faster, but with a sacrifice in extraction efficiency by roughly 5\%. This trade-off becomes more significant at higher pressures.
    \item The difference between the transport efficiency to the entrance versus the end of the neutralization channel illustrates that the main losses occur in the first part of the channel due to the fact that ions (especially peripheral ones) follow the field lines to the channel surface. Diffusion also increases these losses. The problem is most significant at lower pressure, where both diffusion is more significant and the drag force of the gas is weaker. The losses in the channel are less important at higher pressures, where the gas flow can more efficiently guide the ions along the channel axis.
    \item The residence time of the ions in the neutralization channel (time available for neutralization), is nearly equivalent for both the high- and low-voltage settings. This consistency reflects the fact that the ion extraction in the channel only relies on the gas flow, where the gas velocities are independent of the electric field strength. 
    \item At 100~mbar, around 11\% of the particles reach the outlet with an average extraction time of 93~ms for the high-voltage settings and an average residence time in the neutralization channel of 65 ms. At 200~mbar, the extraction efficiency can be more than doubled, with an extraction time of around 126~ms. 
\end{enumerate}

\subsection{Final design and comparison with the current S$^3$ gas cell}

For the final design, the electrode voltages were further optimized to preserve performance, despite the lower number of electrodes reducing the degrees of freedom. Furthermore, adopting a stronger gradient in the funnel and a more abrupt transition to the ground potential of the channel allowed even slightly improving the efficiency, as shown in Table~\ref{tab_eff_t_HV_LV} for the most common working pressures. Ultimately, the applicable voltages will be in practice also be limited by the voltage-breakdown limit of the buffer gas in the cell, which will influence the achievable extraction time. 

\begin{table}
\small 
\centering 
\caption{Simulated performance of the final design with high-voltage settings for 100 and 200~mbar.} 
\vspace*{1mm}
\begin{tabular*}{0.8\linewidth}{@{\extracolsep{\fill}}cccc}
     \hline
     \hline
\makecell{Pressure} &  \makecell{$\epsilon_{\text{ext}}$ (\%)  } & \makecell{$t_{\text{ext}}$ (ms)}& $t_\text{n}$ (ms)  \\
       \hline
\multirow{1}{*}{100 mbar}&14 &93&53\\
  \hline
\multirow{1}{*}{200 mbar}&29 &132&58\\
       \hline
       \hline
 \end{tabular*}
 \label{tab_eff_t_HV_LV}
\end{table}

\begin{figure}
\centering
\includegraphics[width=0.4\textwidth]{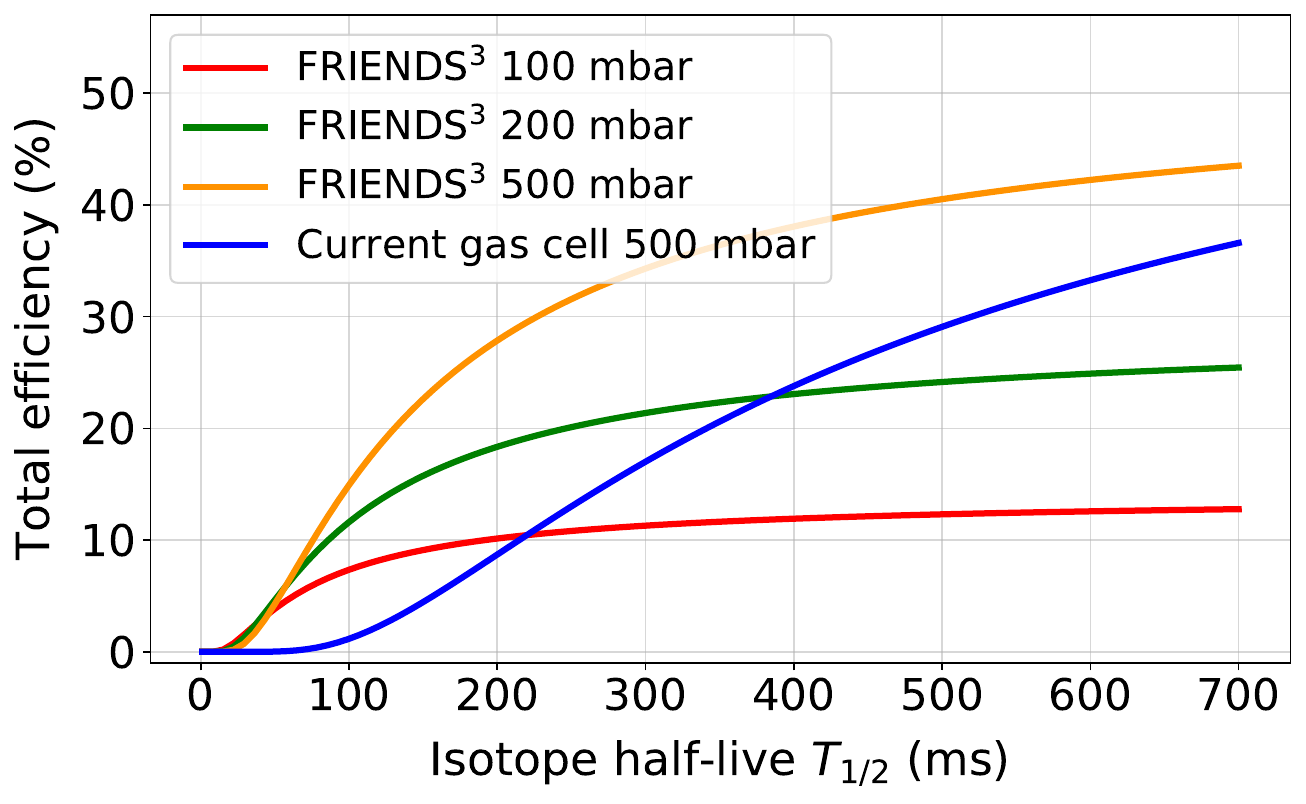}
\caption{Simulated total efficiency of the FRIENDS$^3$ gas cell (considering extraction and decay losses) at 100 mbar, 200 mbar, and 500 mbar, compared to the current gas cell at 500 mbar, for different decay half-lives of the extracted ions.
}
\label{fig:compare_S3_FRIENDS3}
\end{figure}

The performance of the gas cell must account for decay losses when operated with radioactive isotopes. This means that one can define a total extraction efficiency as the product of the transport efficiency up to the exit and the survival fraction $2^{-t_\text{ext}/T_{1/2}}$, which refers to the radioactive ions that do not decay until their extraction from the gas cell. Figure~\ref{fig:compare_S3_FRIENDS3} presents this total efficiency for the final FRIENDS$^3$ gas cell design as a function of the half-life of the transported ions, at various operating pressures: 100 mbar (red), 200 mbar (green), and 500 mbar (orange). The results are  compared to the current gas cell at 500 mbar (blue).  For the S$^3$-LEB gas cell, a simulated extraction efficiency of 65\% was considered and a 580-ms extraction time. These choices will be justified in the next section.

The simulations results indicate that, considering only extraction efficiency and time, the FRIENDS$^3$ gas cell is preferable to the current S$^3$-LEB design for isotopes with short half-lives, particularly those shorter than 200~ms. Additionally, the FRIENDS$^3$ gas cell can be operated at 500-mbar pressure, showing in this case better simulated performance even for half-lives up to 700-800 ms. We note that the neutralization efficiency and the stopping efficiencies of the two gas cells are not considered in the comparison. Having a larger stopping volume and diameter, the FRIENDS$^3$ gas cell should have a stopping efficiency at least as large as the S$^3$-LEB version. The neutralization efficiency of the two designs will be discussed in the next section.

\section{Simulation of neutralization and electron dynamics}
\label{Electron}

\subsection{Transport and neutralization in the S$^3$-LEB gas cell}
\label{sim-S3-LEB}

As mentioned before, upon entering the gas cell, a fraction of the beam will neutralize by direct charge exchange with the argon gas. This fraction has been estimated to be on the order of 10$\%$ \cite{Lautenschlager2016}. While for the S$^3$-LEB gas cell the resulting atoms will be most likely extracted by the gas flow, for the FRIENDS$^3$ version they constitute a loss, because they would be insensitive to the electrical field and their extraction would be very inefficient by the gas flow, due to the very low gas velocity.

For the S$^3$-LEB gas cell, the remaining ions would need to recombine with electrons produced by the ionization of the gas itself, either through three-body or dissociative recombination \cite{kudryavtsev2001gas, Facina2004, Moore2010}. Knowing the linear energy loss of the incident beam in the gas $\Lambda$, the beam flux $\Phi$ and the energy required to produce an electron-ion pair $W$ (26.4~eV for argon), the production rate of argon-ion/electron pairs can be calculated as $p_e = \Lambda \Phi /W$ (in pairs cm$^{-3}$s$^{-1}$). The linear energy loss of the ion beam can be simulated using the Stopping and Range of Ions in Matter (SRIM) simulation package \cite{SRIM2003}, in particular its Transport of Ions in Matter (TRIM) program. 

With the resulting value of $p_e$ for a given beam transversal section and intensity, calculations of the achieved equilibrium density were performed in the past \cite{Ferrer2013}. These calculations,  considering only the electron creation and destruction by argon ionization and recombination Ar $\rightleftharpoons$ Ar$^+$ + $e^-$, are equivalent to taking in Eq.~(\ref{eq:n_e_COMSOL}) only the term $R_e = p_e - \alpha_r n_i n_e$, where $\alpha_r$ is the argon recombination coefficient and $n_i, n_e$ are the ion and electron densities, respectively. Under these conditions, at equilibrium the densities of argon ions and electrons satisfy $n_i=n_e=n_{ie}$ and the time derivative in Eq.~(\ref{eq:n_e_COMSOL}) is zero. The rate equation governing the electron density is then given by:
\begin{equation}
    \frac{dn_{ie}}{dt}=p_e-\alpha_r\cdot n_{ie}^2=0,
\end{equation}
which yields the expression: 
\begin{equation}
n_{ie}=\sqrt{\frac{p_e}{\alpha_r}}
\label{eq:ne}
\end{equation}

For argon at room temperature and 100~mbar, the recombination coefficient 
$\alpha_r$ is 1.0134$\times 10^{-6}$ [cm$^3$/s] \cite{Cooper_van_Sonsbeek_Bhave_1993}, enabling the calculation of $n_{ie}$ as a function of the production rate $p_e$. From $n_{ie}$, the recombination rate for  the stopped ion is simply $R_\text{rec} = \alpha_\text{ion} n_{ie}$, where $\alpha_\text{ion}$ is the total recombination rate of the stopped ion species. 

\begin{figure*}
\centering
\vspace*{2mm}
\includegraphics[width = 0.4\linewidth]{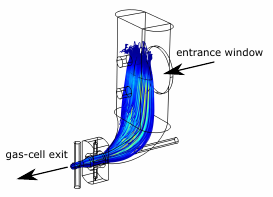}
\includegraphics[width = 0.5\linewidth]{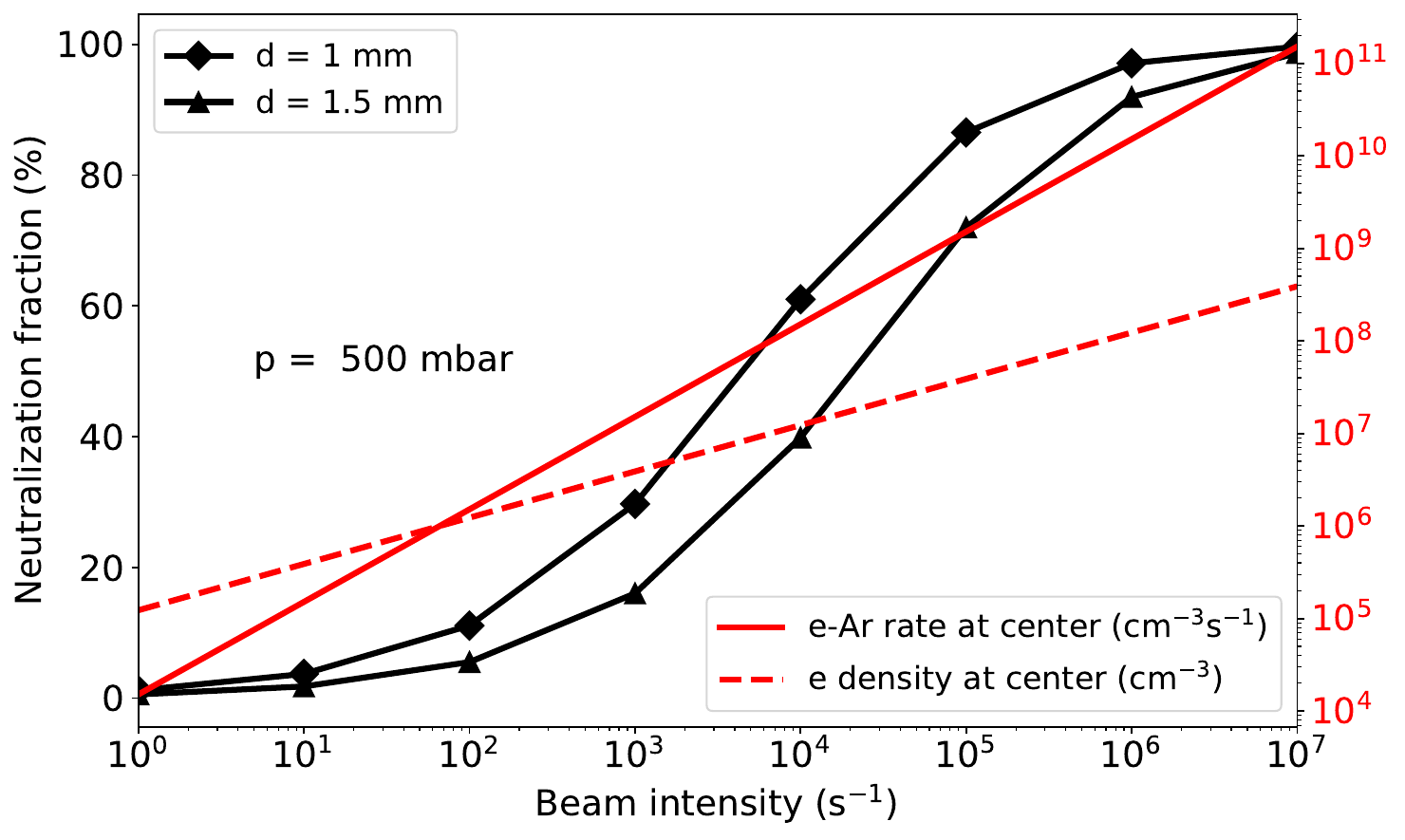}
\caption{(Left) Geometry of the S$^3$-LEB gas cell in the COMSOL simulation space and concentration flow lines resulting from a simulation performed with the COMSOL TDS interface (for 500~mbar stagnation pressure and 1~mm exit-hole diameter). (Right) Simulated neutralization efficiency  (left axis, black) as a function of the beam intensity for a beam of $^{100}$Sn stopped in 500~mbar argon, for exit hole diameters of 1~mm (diamonds) and 1.5~mm (upward triangles). The obtained electron-ion pair production rate (continuous line, in cm$^{-3}$s$^{-1}$) and the equilibrium electron density at the gas cell center (dashed line, in cm$^{-3}$) are also represented in red, corresponding to the right axis. In the approximation leading to Eq.~(\ref{eq:ne}), the two quantities do not depend on the gas-flow velocity.}
\label{S3-LEB-neutr}
\end{figure*}

We have performed a new series of simulations of the S$^3$-LEB gas cell using COMSOL and the TDS physics interface. The 3D gas-cell geometry is shown in the left panel of Fig.~\ref{S3-LEB-neutr}, with the gas inlet defined on the top surface of the volume and the outlet on the exit surface at the lower left side. Such simulations have already been performed in the gas-cell design phase \cite{Kudryavtsev2016}, but have mostly focused on the extraction efficiency and time. Here we include in the simulation the neutralization phenomenon, using the recombination rate based on the equilibrium density calculated with Eq.~(\ref{eq:ne}). We take however into account two aspects: 1) That the linear energy loss of the ions has a distribution with respect to the depth of the gas cell (which we shall refer to as the $z$ coordinate) and the ions are stopped in the tail of this distribution. 2) That the equilibrium ion/electron density is only present in the volume bombarded by the incident ion beam, while outside this volume it should quickly decay to zero (we consider that outside the volume it is zero) due to argon-ion/electron recombination. 

The simulations were performed for an S$^3$ beam of $^{100}$Sn and a 5~$\mu$m titanium entrance window with 500~mbar argon stagnation pressure. To determine some of the input parameters for the COMSOL simulation, we used the TRIM program. A first TRIM simulation was run for $^{100}$Sn arriving perpendicularly on the window, without any spatial spread or divergence. The ions incident on the entrance window were given an average energy of 84.5~MeV and a momentum spread $\Delta p/p = 2.5\%$ as resulting from the transport simulations performed for S$^3$ \cite{Dechery2015}. From this simulation, a discrete sampling of the average linear energy loss per ion $\Lambda$ was extracted as a function of the $z$ coordinate (one value per mm). To determine the spatial parameters of the stopped beam, a second TRIM simulation was performed for $^{100}$Sn ions having not only the average energy and momentum spread from the S$^3$ transport simulations, but also the transversal emittance. The average range of the stopped ions ($z_m$ = 18.5~mm), as well as the longitudinal and radial straggling ($\sigma_l$ = 3~mm and $\sigma_r$ = 9.4~mm, respectively) were used as parameters for the COMSOL simulation. 

The $^{100}$Sn ions were simulated as a diluted species with a time-evolving concentration. To represent the initial ion distribution after thermalization in the stopping volume, a ellipsoidal domain was defined in connection to the TRIM results, having a semiaxis in the $z$ direction equal to $2 \sigma_z$ and a semiaxis in the radial plane equal to $2 \sigma_r$. The domain center was on the symmetry axis of the entrance window at a depth equal to the average range $z_m$ of the stopped ions. 

The gas flow in the cell was simulated using the Laminar Flow physics interface, with the mass flow rate through the inlet adjusted to achieve 500~mbar stagnation pressure.
To simulate the transport of $^{100}$Sn out of the stopping volume, a constant initial concentration was defined within the ellipsoidal domain and a time-dependent simulation was performed for the TDS interface, computing the total flux through the exit surface of the gas cell. The critical parameter of the simulation was the diffusion constant of $^{100}$Sn in argon, which was estimated to be 1.6$\times 10^{-5}$ m$^2$/s (at 500~mbar pressure) using the Chapman–Enskog formula \cite{Chapman1970}. By analyzing the time dependence of the output flux, the extraction time of the gas cell could be determined. The transport efficiency was calculated as the ratio between the time integral of the total output flux and the concentration integral on the start ellipsoidal domain. The simulations were performed for exit-hole diameters of 1~mm and 1.5~mm. The former led to an extraction time of 580 ms and extraction efficiency of 65\%, the latter led to an extraction time of 280~ms and an efficiency close to 100~\%. The concentration flow lines resulting from a simulation performed for an exit-hole diameter of 1~mm are shown in the left panel of Fig.~\ref{S3-LEB-neutr}.

To simulate the effect of neutralization, the discrete linear energy loss per ion $\Lambda$ extracted from TRIM was fitted with a fifth order polynomial, allowing to express it analytically as a function of the $z$ coordinate. The resulting function was implemented in COMSOL and, for a given incident beam intensity, it was used to calculate the ion/electron pair production rate and from it the equilibrium density at any position in space. In the calculation, the beam cross-section was considered circular and uniform, with a radius equal to $2 \sigma_r$.  The value of the equilibrium density was then used to calculate a space-dependent recombination rate for the stopped ions. To simplify the implementation, the resulting recombination rate was considered radially constant (for a given depth $z$) within the cross section of the incident beam and zero outside. The effect of recombination was then implemented as a reaction rate with the described spatial distribution, diminishing the concentration of the transported species.

The simulation with the TDS interface was then performed with the recombination rate active, for different intensities of the incident beam, which as described before scale proportionally the ion-electron pair production rate. For each intensity, the time-integrated output flux was computed and the difference between the result without/with recombination was interpreted as the neutralized substance. Taking the ratio between this quantity and the integral of the initial concentration over the ellipsoidal start domain was  the neutralization efficiency. The results are presented in the right panel of Fig.~\ref{S3-LEB-neutr} as a function of beam intensity, where the left-side $Y$ axis represents the neutralization efficiency and the right-side $Y$ axis is shared between the ion-electron production rate at the gas-cell center (in cm$^{-3}$ s$^{-1}$) and the equilibrium electron density at the same position (in cm$^{-3}$). The neutralization-efficiency results are presented for the two considered exit-hole diameters. One notices that, despite the restriction of the recombination to the volume bombarded by the incident beam, a significant fraction of the beam (above 10\%) is expected to be neutralized for beam intensities above $10^3$~ions/s. Nevertheless, for intensities below $10^2$~ions/s, it is expected that the recombination will be smaller compared to initial neutralization fraction resulting from direct charge-exchange in the stopping phase. Furthermore, increasing the exit-hole diameter can significantly reduce the recombination efficiency due to the faster extraction of the ions from the stopping volume.

\subsection{Electron dynamics in the neutralization channel of the FRIENDS$^3$ setup}
\label{sim-electrons}

In the simulations of the previous subsection, the electrons and ions produced by the ionization of the argon gas by the ion beam were considered static, while the evolution of their densities was only determined by their production $p_e$ and  recombination rate $\alpha_{r} n_{e} n_{i}$, leading to the equilibrium result of Eq.~(\ref{eq:ne}). Nevertheless, the considerable diffusion of the electrons, their convection in the gas flow and their migration in the electrical field lead to the additional terms of Eq.~(\ref{eq:n_e_COMSOL}) which can have a sizable impact. We have thus performed a simulation of the electron-density evolution following the ionization of the gas, by taking into account all these effects. 

As a case study, we have used the neutralization channel of the FRIENDS$^3$ gas-cell design (see Fig.~\ref{fig:geo}). The simulation is performed in cylindrical symmetry and the geometry models only the neutralization channel, corresponding to a tube of internal diameter of 18~mm and length 60~mm, with a final converging section identical to the one used in the gas-cell simulations. 
\begin{figure}[ht!]
\centering
\includegraphics[width = \linewidth]{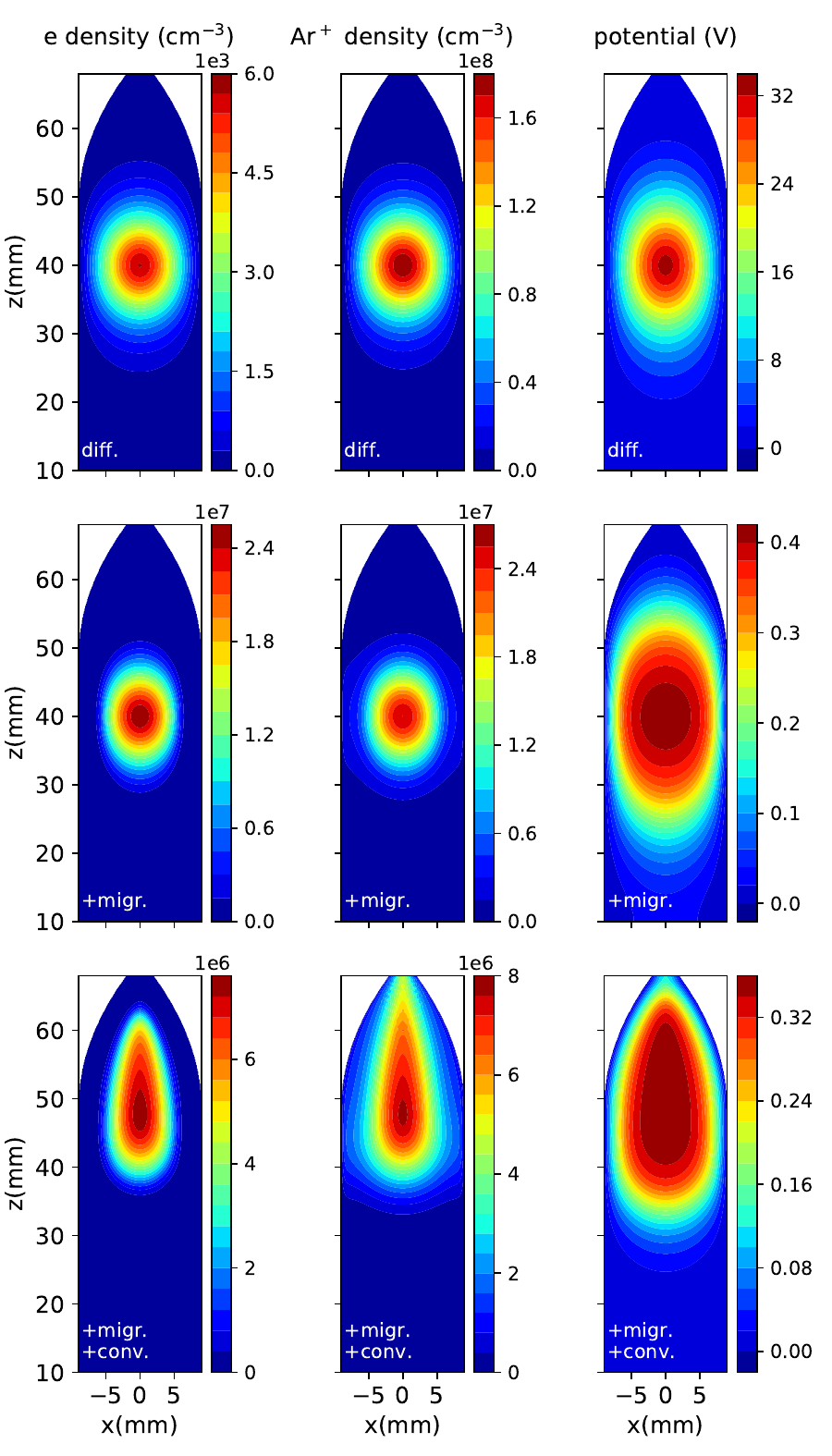}
\caption{Representative simulation results of equilibrium densities following the ionization of the argon buffer gas at 100~mbar pressure in the neutralization channel of the FRIENDS$^3$ prototype gas cell: (left column) electron density, (center column) ion density, (right column) electrical potential. The different rows represent the results with different effects taken into account, namely diffusion only (top), diffusion plus migration (middle), diffusion plus migration and convection (bottom).  All results correspond to an electron-ion pair production rate with a Gaussian profile centered at $x = 0$ and $z=40$~mm, having z/x standard deviations of 4/3.5~mm and a peak value of $10^9$~cm$^{-3}$s$^{-1}$.}
\label{fig:densities}
\end{figure}
The gas flow is modeled identically to the COMSOL simulations presented in the previous sections.
The Plasma module of COMSOL was used to study both the electron and ion dynamics. The production of electron-ion pairs was modeled without any description of the underlying mechanism, by a reversible reaction Ar $\rightleftharpoons$ Ar$^+ + e^-$, with the direct rate given as an input parameter of the simulation, and the reverse rate defined by the argon recombination coefficient $\alpha_r$. For ensuring a smooth convergence of the simulation, the direct rate was modeled with a Gaussian spatial distribution:  
\begin{equation}
    k^f=p_e \cdot \exp \left[\frac{-(z-z_0)^2}{2\sigma_z^2}\right]\cdot \exp \left[\frac{-r^2}{2\sigma_r^2}\right],
    \label{kf}
\end{equation}
where $z_0$ represents the position at which the ionization rate is $p_e$ (in this work 30~mm from the beginning of the tube), while $\sigma_z$ and $\sigma_r$ are the axial and radial standard deviations of the ionization volume (in this work 4~mm and 3.5~mm respectively).

In the simulations, no inelastic scattering processes between the electrons and the argon atoms were considered. The electrons and argon ions could only vanish by recombination, or on the boundaries of the domain, without any secondary-electron emission. The domain boundaries were considered at ground potential and no external electrical field was taken into account. 

The critical parameters of the simulation are the mobilities and diffusion coefficients in argon of the electrons and argon ions. The electron mobility varies exponentially with the reduced electrical field in the low-field regime. Because in the equilibrium solution the plasma potential was found to be lower than 0.5 V, we have taken for the electron mobility a value of 40420~cm$^2$/(V s), which corresponds to a reduced electrical field of 0.01~Td (below this value the mobility reaches a plateau) \cite{Pack1961}. The experimental data on the electron diffusion constant shows a dip for reduced electrical fields of $\approx 3$~Td, with a gradual increase towards lower field values, but showing overall less variability \cite{Nakamura1988}. For this quantity we take an average of 1300~cm$^2$/s describing the overall range of experimental values. Based on a calculation from \cite{Pack1992} it is also possible to estimate the diffusion constants for a reduced field of 0.01~Td to be 3234~cm$^2$/s. Both mobility and diffusion constant are given for 1~atm pressure and 293.15~K temperature. In the simulations, we have taken into account this uncertainty in the electron diffusion constant by computing the results for both values.  

For argon ions the transport properties depend much less on the electrical-field value in the medium-low field regime, therefore we use the low-field reduced mobility of 1.53 cm$^2$/(V s) \cite{Ellis1976}, while the diffusion constant was calculated using the Einstein relation to 3.9$\times 10^{-2}$cm$^2$/s (both defined for 1~atm and 273.15~K). 

The simulation with the Plasma module can only be time dependent, therefore for this study particular attention was given to the numerical stability of the simulation. Plasma simulations deal with particle densities which can vary by several orders of magnitude over the simulation space and approach zero in some areas, which can lead to instabilities. Therefore, test simulations were performed to ensure that the time stepping conserves the initial particle densities in the absence of any processes. For some of the simulations, a stabilization "source term" was added, which is a term varying as a power $\zeta$ of the inverse of the electron density. This standard COMSOL procedure \cite{COMSOL5.6_plasma} helps eliminate divergencies associated to small density values. When used, several values of the $\zeta$ parameter were tested and, in case of variation, all results were represented. For all studies, the time evolution of the system was simulated for $10^5$~s, to facilitate the detection of all numeric instabilities. 

\begin{figure}[ht!]
\centering
\vspace*{2mm}
\includegraphics[width = \linewidth]{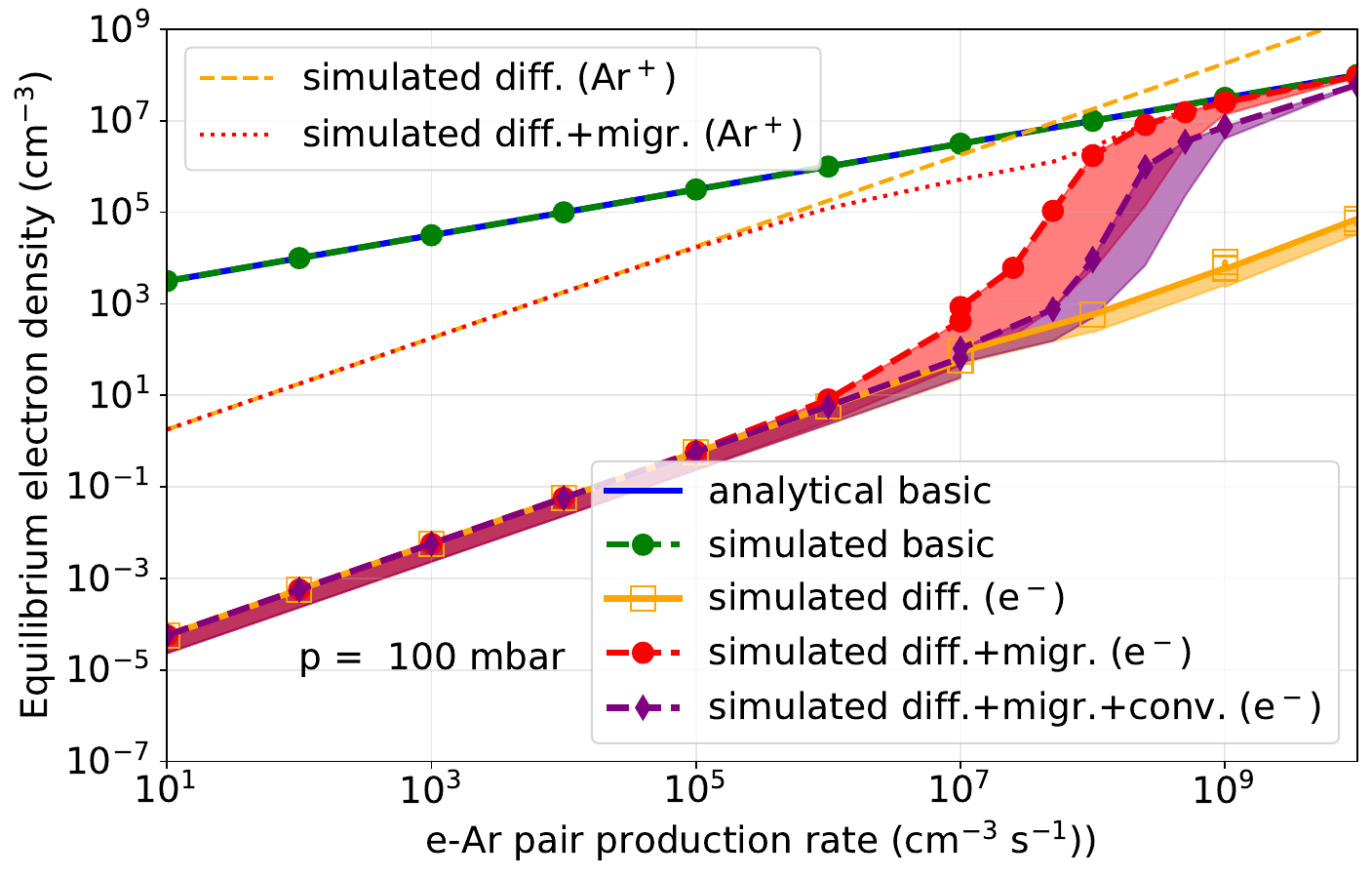}
\includegraphics[width = \linewidth]{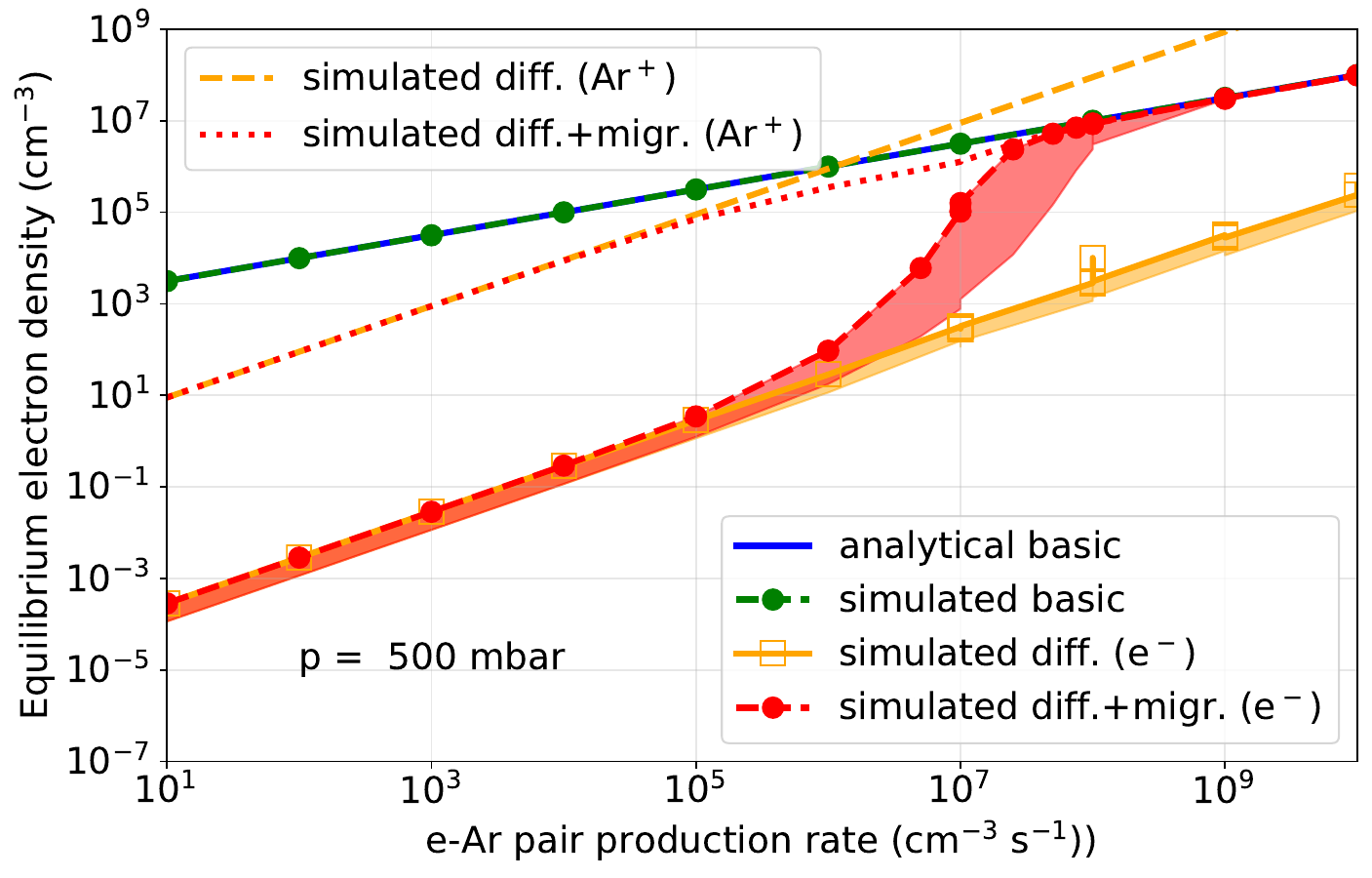}
\caption{Simulated equilibrium electron/ion densities in the neutralization channel of the gas-cell prototype (at the peak of the distribution, see Fig.~\ref{fig:densities}) as a function of the production rate (also at the peak) for different effects included in the simulation. The gas pressure is 100~mbar (top panel) and 500~mbar (bottom panel). The results of the basic analytical result of Eq.~(\ref{eq:ne}) are represented by a blue line, compared to the simulations performed in the same conditions (full green circles connected by dashed line). The results with only diffusion taken into account are shown in orange (empty squares connected by continuous line for electrons, simple dashed line for ions). The results with diffusion and migration in the electrical field are shown in red (full circles connected by dashed line for electrons, simple dotted line for ions). Finally, the electron density considering diffusion, migration and convection in the gas flow is represented with purple diamonds connected by a dashed line (for 100~mbar only). The data points for electrons are for a value of the diffusion constant of 1300~cm$^2$/s. Below the data points, a shaded area shows the difference to the results obtained with a diffusion constant of 3234~cm$^2$/s. The results for ions are only presented for the lower of the two electron diffusion constants.}
\label{fig:comsol-python-ne}
\end{figure}

With this approach, a systematic series of simulations was performed with the different terms of the electron/ion dynamics gradually switched on. The spatial electron/ion densities of some representative cases at equilibrium are shown in Fig.~\ref{fig:densities} (electron density on the left column, argon ion density on the middle column and electrical potential on the right column). The results of all the studies are presented in Fig.~\ref{fig:comsol-python-ne} as a function of the electron/argon-ion pair production rate. For all simulations, the resulting densities have the profile of a single peaked distribution. The density values presented in Fig.~\ref{fig:comsol-python-ne} correspond to the apex of these distributions. Furthermore, due to the uncertainty in the electron diffusion constant, the data points for electrons in Fig.~\ref{fig:comsol-python-ne} correspond to a diffusion constant of 1300~cm$^2$/s, while the shaded areas under the data points present the difference to the (lower) results obtained for a value of the diffusion constant of 3234~cm$^2$/s. The results presented for ions are only for an electron diffusion constant of 1300~cm$^2$/s. Similarly, the densities presented in Fig.~\ref{fig:densities} correspond to this value. 

Before studying the effect of the terms not considered in the previous subsection, a first series of simulations was performed in order to check whether COMSOL can correctly describe the basic case in which only the ionization and recombination are active, which should lead to a result described by Eq.~(\ref{eq:ne}). This is seen to be true in  Fig.~\ref{fig:comsol-python-ne}, with the top panel corresponding to 100~mbar and the lower panel to 500~mbar stagnation pressure. Indeed, the continuous blue line represents the analytical result of Eq.~(\ref{eq:ne}), while the green dots connected by a dashed line show the simulated results. The two sets of results are in perfect agreement. We note that for this particular set of simulations, all wall effects were removed, in order to avoid having an additional loss channel for the charged particles.

In a second set of simulations, represented in Fig.~\ref{fig:comsol-python-ne} by the orange squares (electron density) and orange dashed line (argon-ion density) the diffusion of argon ions and electrons is switched on. As one can see, the very large diffusion constant of electrons determine a dramatic drop in their equilibrium density, with the ions at a systematically larger value (due to their lower diffusion constant), albeit lower than the result of Eq.~(\ref{eq:ne}) for production rates below $10^7$~cm$^{-3}$~s$^{-1}$. A result obtained under these conditions for a production rate of $10^9$~cm$^{-3}$~s$^{-1}$ is shown in the first row of Fig.~\ref{fig:densities}. One can notice the very large gap between the argon and electron density, which leads to a considerable positive potential distribution in the production area (as high as 30~V). This potential only becomes non-negligible for argon-ion densities above $10^6$~cm$^{-3}$. 

By activating the migration in the electrical field for both electrons and ions, the effect of the electrical potential resulting from the charge densities becomes important. For low production rates, in which the electrical potential is negligible despite the large asymmetry between the argon and electron densities, there is no observable effect. Thus, the resulting densities, represented in red circles for the electrons and red dots for the ions, follow closely the results obtained with only diffusion being considered. 

Nevertheless, for large enough production rates, the significant electrical potential developing due to the argon-ion density at the center of the neutralization channel acts as a well for the electrons, compensating (more and more as the production rate increases) the effect of diffusion. One thus observes a gradual transition from the diffusion-dominated, non-neutral regime to one where the electron and ion densities follow very closely the values obtained with Eq.~(\ref{eq:ne}). In this regime, represented in the second row of Fig.~\ref{fig:densities} for a production rate of $10^9$~cm$^{-3}$~s$^{-1}$, the positive and negative charges are almost perfectly balanced, leading to a much shallower potential well for the electrons. For large-enough production rates, the significantly lower diffusion constant of the argon ions leads, via their attractive potential, to a stabilization of the equilibrium electron densities around the values obtained without diffusion (dynamically suppressing the diffusion effect on the electrons). This result is similar for 100~mbar and 500~mbar stagnation pressure, the only difference being a transition to the equilibrium regime at slightly lower production rate, due to the lower diffusion constant (at the higher pressure) of both the ions and electrons.   

In a final set of simulations, the effect of convection in the gas flow was also enabled and the resulting electron densities are presented in the top panel of Fig.~\ref{fig:comsol-python-ne} by the purple diamonds (only for 100~mbar). The spatial distributions are presented in the bottom row of Fig.~\ref{fig:densities} for a production rate of $10^9$~cm$^{-3}$~s$^{-1}$. The gas flow will significantly interact with the argon ions, since the drag force is inversely proportional to the mobility and will mostly impact ions. This moves the ion distribution downstream, distorts it along the gas-flow lines and modifies accordingly the resulting potential well. This in turn attracts the electron distribution downstream and gives it a similar distorted shape. The flow moves the equilibrium densities away from the electron-ion-pair production region, which leads to somewhat lower densities, although they remain very close to the basic result of Eq.~(\ref{eq:ne}). The transition to the charge-equilibrium regime also occurs for larger production rates than in the absence of the gas flow. 

The general conclusion of these simulations is that the electron-diffusion phenomenon needs to be taken into account when estimating the achievable density of the electron cloud from the ionization of the buffer gas, especially for low and medium ionization rates and especially at lower pressures. Convection in the gas flow also needs to be accounted for when the electron cloud is produced close to the gas-cell exit, as is the case for the FRIENDS$^3$ gas cell. If one uses the larger value of the diffusion constant for electrons (see shaded area in Fig.~\ref{fig:comsol-python-ne}), one notices that the results change especially around the transition point to the equilibrium regime, which shifts to $\approx 4-5$ times larger electron-ion pair production rate. This transitional area is also the only one where there is a significant variation (not shown) in the obtained ion densities, but by less than a factor 3, which does not change significantly the picture drawn in Fig.~\ref{fig:comsol-python-ne}.

Although the simulations were not performed also for the exact geometry of the S$^3$-LEB gas cell, a qualitative discussion can be made using the results obtained for 500~mbar in the neutralization channel. In the stopping volume of the S$^3$-LEB gas cell, the gas flow is negligible, so the effect of convection is not important. Diffusion will however play a role, and based on the results of our study it appears that electron-ion production rates of at least $10^8$~cm$^{-3}$~s$^{-1}$ are needed to certainly approach the equilibrium densities for which significant recombination can occur (and conversely, that negligible recombination can be expected for rates below $10^7$~cm$^{-3}$~s$^{-1}$). In between, the conclusion strongly depends on the electron diffusion constant. This means that ion-beam intensities larger than a few $10^3$s$^{-1}$ are required. Compared to Fig.~\ref{S3-LEB-neutr}, one can see that these are also the beam intensities for which (even in the basic case) a significant neutralization can be expected. This is the consequence of the very efficient ionization of the argon gas by the heavy-ion beam. 

For the cases in which the stopped beam would have insufficient intensity to provide the necessary electron-ion pair production rate, the use of a beta source was proposed as a means to ionize the buffer gas in the neutralization volume. For a gas-cell design similar to Fig.~\ref{fig:geo}, the beta source could be coupled laterally (for example through a thin window) to the neutralization channel. A pure-beta source like $^{90}$Sr would also avoid any radioprotection concerns. 

However, the ionization of the argon by energetic betas is not very efficient, especially at lower pressures. By using energy-loss data for betas in argon and the energy spectrum of a $^{90}$Sr beta source, it is possible to estimate the required activity for reaching a certain electron-ion pair production rate. In order to achieve production rates on the order of $10^8$~cm$^{-3}$~s$^{-1}$, a $^{90}$Sr beta source with a diameter of 15~mm (commercially available) would need to have activities (delivered to the gas) on the order of 10~MBq. For example, taking an activity of 30~MBq for a beta source with 180$^\circ$ divergence and assuming that it can be brought as close as 15~mm from the neutralization channel and coupled through a 15~mm-diameter window (without any attenuation), the resulting ionization rate at the center of the neutralization channel would be $6 \times 10^7$ cm$^{-3}$ s$^{-1}$ for 100~mbar pressure and $3 \times 10^8$ cm$^{-3}$ s$^{-1}$ for 500~mbar.

The value for 500~mbar, neglecting any gas-flow effect, is in the equilibrium regime (see bottom panel of Fig.~\ref{fig:comsol-python-ne}) and would be such if the beta source were coupled in the stopping volume of the S$^3$-LEB gas cell. In this case, one can read off of Fig.~\ref{S3-LEB-neutr} that the corresponding electron density would lead to a neutralization efficiency above 50\% for an exit-hole diameter of 1~mm. 

In the neutralization channel of the FRIENDS$^3$ gas cell, the gas velocity is around 1~m/s and thus the ions require 15~ms to cross the volume in front of the beta source (which can be considered the effective neutralization time). Not considering the effect of the diffusion or gas flow  on the electron density, the electron-ion pair production rate would lead to a recombination lifetime of about 125~ms for 500~mbar, which means a neutralization efficiency in 15~ms of about 10\%. However, for 100~mbar, one can observe in the top panel of Fig.~\ref{fig:comsol-python-ne} that diffusion and gas flow put the rate of $6 \times 10^7$ cm$^{-3}$ s$^{-1}$ in the diffusion-dominated regime, leading to a dramatic reduction of the electron density. Consequently, using a beta source for neutralization in 100~mbar with the simulated neutralization-channel geometry appears unfeasible. Nevertheless, operating the gas cell at higher pressure (200-300~mbar) and using a longer neutralization channel, to avoid the electron production occurring too close to the outlet, or potentially a more efficient overlap of the ion trajectories with the emission cone of the beta source, could give rise to a usable configuration, providing acceptable neutralization efficiencies. An experimental setup allowing to study the effect of a $\approx$30~MBq $^{90}$Sr source is currently being investigated in connection to the FRIENDS$^3$ experimental setup.

\section{Conclusions}

In this work, we have presented a series of simulations of gas stopping cells aimed at producing neutral beams of radioactive isotopes for laser ionization and spectroscopy, in the context of the S$^3$-LEB installation at SPIRAL2-S$^3$. We simulated an alternative design of the S$^3$-LEB gas cell compared to the version currently installed at the focal plane of the S$^3$ spectrometer, aimed at reducing the extraction time of the ions from the stopping volume by the use of an electrical field. Different simulation methods using COMSOL and SIMION were described in detail and applied to the simulation of a conceptual design, giving comparable results. The final, optimized prototype gives an improved total efficiency compared to the current S$^3$-LEB gas cell for ions of half-life below 500~ms if one considers the combined losses due to transport and decay during the extraction time. 

A second part of this work was dedicated to a study of the neutralization of the ion beam by recombination with electrons produced by the ionization of the buffer gas. The simulations of the S$^3$-LEB gas cell were updated, taking into account the spatial distribution of the equilibrium electron density produced by the stopped ion beam and the extension of the volume where the neutralization can occur. We show that for the standard gas-cell operation conditions, a neutralization efficiency of the ion beam larger than 50\% is simulated for incident ion rates above $10^4$~ions/s.  

Using a geometry similar to the neutralization channel of the new gas-cell prototype, a series of simulations was performed with the Plasma module of COMSOL, aimed at quantifying the effect of diffusion, migration and convection on the achieved electron and ion densities. We show that for low electron-ion production rates the electron dynamics is dominated by diffusion, leading to a collapse of electron density. Beyond a critical production rate, a transition to a charge-equilibrium regime occurs due to the attractive potential created by the argon ions, where the densities follow closely the analytical values obtained when diffusion is neglected. We show that this regime can be reached in the stopping volume of the current S$^3$-LEB gas cell. In the case of ionization in the neutralization channel of the new prototype by a beta source of $\approx 30$~MBq at 100~mbar pressure, insufficient electron-ion pair production and the effect of diffusion/gas flow put the system in a diffusion-dominated regime. 

The gas cell prototype was manufactured and an experimental setup was recently assembled for characterizing it \cite{Morin2025}. A series of experimental tests will be performed in order to verify the simulation results, quantify the performance of the new gas-cell prototype, and study the effect of gas-ionization by a beta source and other neutralization mechanisms in an optimized configuration. Further simulations will also be performed to study the effect of the chamber geometry or of the ionization-rate spatial distribution on the results of this work. The impact of space charge on the extraction efficiency of the gas cell under the action of electrical fields will also be studied in future work. 

\section*{CRediT authorship contribution statement} 
\textbf{W.~Dong:} Conceptualization, Methodology, Investigation, Formal analysis, Visualization, Writing - Original Draft. \textbf{V.~Manea:} Conceptualization, Methodology, Investigation, Supervision, Funding acquisition, Writing - Original Draft. \textbf{R.~Ferrer:} Conceptualization, Methodology, Investigation, Resources \textbf{S.~Franchoo:} Methodology, Resources, Writing - Review  \& Editing. \textbf{S.~Geldhof:} Methodology, Resources, Writing - Review  \& Editing. \textbf{F.~Ivandikov:} Methodology, Investigation, Writing - Review  \& Editing. \textbf{N.~Lecesne:} Methodology, Resources. \textbf{D.~Lunney:} Methodology, Investigation, Resources, Writing - Review  \& Editing. \textbf{V.~Marchand:} Investigation. \textbf{E.~Minaya Ramirez:} Methodology, Resources \textbf{E.~Morin:} Methodology, Investigation, \textbf{S.~Raeder:} Conceptualization, Methodology, Writing - Review  \& Editing. 

\section*{Declaration of competing interest}

The authors declare that they have no known competing financial interests or personal relationships that could have appeared to influence the work reported in this paper.

\section*{Acknowledgments}

This work was funded by the French Research Ministry through the National Research Agency under contract number ANR-21-CE31-0001, by the French IN2P3 and GSI under the French-German collaboration agreement number PN1064, by European Union’s Horizon 2020 research and innovation programme under Grant Agreement No. 861198-LISA-H2020-MSCA- ITN-2019, by the Research Foundation Flanders (FWO, Belgium) BOF KU Leuven (C14/22/104) and by the FWO under the Excellence of Science (EOS) program (40007501). The co-authors would like to thank the S$^3$-LEB collaboration for fruitful exchanges on the topics related to the FRIENDS$^3$ project.

\bibliographystyle{elsarticle-num} 
\bibliography{Biblio} 

\end{document}